\pdfoutput=1
\documentclass{iopjournal}

\usepackage[T1]{fontenc}
\usepackage{lmodern}
\usepackage{cite}
\usepackage{amsmath,amssymb,amsfonts}
\usepackage{amsthm}
\usepackage{braket}
\usepackage{algorithmic}
\usepackage{algorithm}
\usepackage{booktabs}
\usepackage{array}
\usepackage{multirow}
\usepackage{tikz}
\usetikzlibrary{arrows.meta,positioning,fit,backgrounds,calc}

\newcommand{\rev}[1]{#1}
\newenvironment{revised}{\par}{\par}

\newcommand{\NUMrounds}{85}
\newcommand{\NUMdparams}{42}
\newcommand{\NUMnqubits}{6}
\newcommand{\NUMnlayers}{3}
\newcommand{\NUMlr}{0.05}
\newcommand{\NUMwindow}{10}
\newcommand{\NUMepsstop}{0.01}
\newcommand{\NUMmaxiter}{150}
\newcommand{\NUMrunsmain}{200}
\newcommand{\NUMrunsother}{100}
\newcommand{\NUMTlo}{30}
\newcommand{\NUMThi}{54}
\newcommand{\NUMghi}{1.8}
\newcommand{\NUMTspread}{24}
\newcommand{\NUMroundspread}{2070}

\newcommand{\NUMaccTolOne}{76\%}
\newcommand{\NUMaccMax}{73\%}
\newcommand{\NUMaccTolMax}{91\%}
\newcommand{\NUMnsessionsMax}{30}

\newcommand{\NUMmeanTEx}{44}

\newcommand{\NUMTmaxEx}{150}

\newcommand{\NUMnlSlope}{0.95}
\newcommand{\NUMnlRsq}{0.995}
\newcommand{\NUMnlDev}{3.1}
\newcommand{\NUMhessD}{42}

\newcommand{\NUMspsaTlo}{47}
\newcommand{\NUMspsaThi}{89}

\newcommand{\RVdecOne}{48\%}
\newcommand{\RVdecTwo}{3.1\%}

\newcommand{\RVearlyR}{0.990}
\newcommand{\RVearlyRho}{0.985}
\newcommand{\RVearlyLoo}{1.7}

\newcommand{\RVgLoo}{3.6}
\newcommand{\RVstopLoo}{3.8}

\newcommand{\RVnullLoo}{10.3}

\newcommand{\RVmuEarlyRatio}{2}
\newcommand{\RVwithinStop}{-0.56}
\newcommand{\RVwithinEarly}{+0.87}

\newcommand{\RVmaxI}{1.68}
\newcommand{\RVminSigI}{0.14}
\newcommand{\RVmainI}{1.24}
\newcommand{\RVmainSE}{0.019}
\newcommand{\RVmainBA}{1.10}
\newcommand{\RVindI}{0.43}
\newcommand{\RVindBA}{0.23}
\newcommand{\RVindSNR}{1.02}
\newcommand{\RVmainSNR}{2.74}
\newcommand{\RVnFourI}{0.53}
\newcommand{\RVnEightI}{0.81}

\newcommand{\RVspsaI}{0.30}
\newcommand{\RVspsaSNR}{1.04}
\newcommand{\RVqsI}{0.007}
\newcommand{\RVqsP}{0.39}
\newcommand{\RVqgI}{0.14}
\newcommand{\RVqgP}{0.005}

\newcommand{\RVqgBound}{0.17}
\newcommand{\RVqpI}{-0.002}
\newcommand{\RVqpP}{0.50}
\newcommand{\RVxfWorst}{78\%}
\newcommand{\RVowOneLo}{0.76}
\newcommand{\RVowOneMed}{0.84}
\newcommand{\RVowOneHi}{0.97}
\newcommand{\RVowSixLo}{0.85}
\newcommand{\RVowSixMed}{0.95}
\newcommand{\RVowSixHi}{1.00}
\newcommand{\RVtprMed}{0.27}
\newcommand{\RVtprLo}{0.03}
\newcommand{\RVtprHi}{1.00}
\newcommand{\RVxsDiag}{0.34}
\newcommand{\RVxsOffLo}{0.24}
\newcommand{\RVxsOffHi}{0.53}

\newcommand{\RVxrOffMean}{0.08}
\newcommand{\RVxrOffMax}{0.15}
\newcommand{\RVtpMaxDiff}{2.3}
\newcommand{\RVskewLo}{0.14}
\newcommand{\RVskewHi}{0.51}
\newcommand{\RVskewLC}{-0.39}
\newcommand{\RVlfReject}{2}
\newcommand{\RVprocLo}{28\%}

\newcommand{\RVprocHi}{79\%}
\newcommand{\RVbudFiftyB}{44}
\newcommand{\RVbudFiftyTrunc}{50\%}
\newcommand{\RVbudFiftydE}{0.032}
\newcommand{\RVbudFiftyRounds}{+0.2}
\newcommand{\RVbudSevFiveB}{52}
\newcommand{\RVbudSevFivedE}{0.0065}

\newcommand{\RVindBudFiftydE}{0.034}
\newcommand{\RVsvtSixFourI}{1.07}

\newcommand{\RVsvtSixteenI}{0.11}
\newcommand{\RVsvtSixteenEarly}{92\%}
\newcommand{\RVsvtSixteendE}{0.23}
\newcommand{\RVsvtEightEarly}{96\%}
\newcommand{\RVsvtEightdE}{0.34}
\newcommand{\RVsvtLBEight}{49\%}
\newcommand{\RVsvtLBSixteen}{38\%}
\newcommand{\RVunprotRounds}{43.8}
\newcommand{\RVlamMinLoo}{11.4}
\newcommand{\RVlamMinLooB}{13.7}
\newcommand{\RVlamMinLo}{0.012}
\newcommand{\RVlamMinHi}{0.022}
\newcommand{\RVtauSlowLo}{465}
\newcommand{\RVtauSlowHi}{855}
\newcommand{\RVrankLo}{22}
\newcommand{\RVrankHi}{33}
\newcommand{\RVdecOneRatio}{3.9\%}
\newcommand{\RVnCfgU}{14}
\newcommand{\RVnSigU}{12}
\newcommand{\RVxfRatioLo}{0.39}
\newcommand{\RVxfRatioHi}{0.84}
\newcommand{\RVxfWorstName}{Max-Cut density under gradient descent}
\newcommand{\RVtpdt}{4.6}
\newcommand{\RVtpsig}{4.7}
\newcommand{\RVtpN}{12}
\newcommand{\RVtpEmpSixteen}{4.8\%}
\newcommand{\RVtpThSixteen}{2.5\%}
\newcommand{\RVaccMain}{34\%}
\newcommand{\RVindAcc}{19\%}
\newcommand{\RVresGap}{0.30}
\newcommand{\RVbudFiftyPct}{10\%}
\newcommand{\RVbudSevFivePct}{2\%}
\newcommand{\RVsvtSixteenPct}{76\%}
\newcommand{\RVpadFiveI}{0.31}
\newcommand{\RVbudSevFiveExtra}{8.2}
\newcommand{\RVbudNineNineExtra}{20.2}
\newcommand{\RVpadFiveExtra}{18.4}

\newcommand{\RVprocSdLo}{2.2}
\newcommand{\RVprocSdHi}{2.7}
\newcommand{\RVowWorstG}{2.0}

\newcommand{\RRriseMain}{0.6\%}

\newcommand{\RRtruncMain}{3128}
\newcommand{\RRriseInd}{1.3\%}

\newcommand{\RRtruncInd}{1513}

\newtheoremstyle{ieeeplain}{0.6em}{0.6em}{\itshape}{}{\bfseries}{.}{ }{}
\theoremstyle{ieeeplain}
\newtheorem{theorem}{Theorem}
\newtheorem{lemma}[theorem]{Lemma}
\newtheorem{corollary}[theorem]{Corollary}
\newtheorem{proposition}[theorem]{Proposition}

\newtheoremstyle{ieeedef}{0.6em}{0.6em}{}{}{\bfseries}{.}{ }{}
\theoremstyle{ieeedef}
\newtheorem{definition}{Definition}
\newtheorem{assumption}{Assumption}
\newtheorem{remark}{Remark}
\newtheorem{hypothesis}{Hypothesis}

\newcommand{\Leak}{\mathcal{L}}
\newcommand{\Ex}{\mathbb{E}}
\let\Pr\relax
\DeclareMathOperator*{\Pr}{\mathbb{P}}
\newcommand{\Var}{\mathrm{Var}}
\newcommand{\Iinfo}{\mathrm{I}}
\newcommand{\dTV}{\delta_{\mathrm{TV}}}
\newcommand{\Nat}{\mathbb{N}}
\newcommand{\eps}{\varepsilon}
\newcommand{\SNR}{\mathrm{SNR}}
\newcommand{\Lap}{\mathrm{Lap}}

\begin{document}

\title{\rev{Stopping-Time Leakage in Blind Quantum Delegation}}

\author{Noureldin Mohamed$^{1,*}$\orcid{0009-0001-4150-8690} and Saif Al-Kuwari$^{1}$\orcid{0000-0002-4402-7710}}

\affil{$^1$Qatar Center for Quantum Computing, College of Science and Engineering, Hamad Bin Khalifa University, Doha, Qatar}

\affil{$^*$Author to whom any correspondence should be addressed.}

\email{nomo89098@hbku.edu.qa}


\begin{abstract}
Blind quantum computation lets a client delegate a computation while
hiding it from the server, up to a declared leakage. Its security
definitions cover a single round. A variational quantum algorithm,
however, delegates many rounds, and its classical optimizer decides from
private data when to stop. We study what the resulting round count
tells an honest-but-curious server. For protocols with one-time-padded
transcripts and fresh randomness in every round, we prove that the
server's best distinguishing advantage over the whole sequence equals
the total-variation distance between the stopping-time distributions,
so the round count is the only residual channel in this model. We also
show that soundness composes as $\delta\,\Ex[T]$ under adaptive
stopping\rev{, given a conditional per-round soundness property that
composably secure verifiable protocols provide}. In simulated delegated eigensolvers for the transverse-field
Ising chain, the local Polyak--{\L}ojasiewicz ratio measured in iterations
5 to 15 predicts the noiseless stopping round with a leave-one-out
error of \RVearlyLoo{} iterations, compared with \RVnullLoo{} for a
constant predictor. Across \RVnCfgU{} configurations that differ in
optimizer, stopping rule, initialization, system size, shot budget and
problem family, the leakage is significant in \RVnSigU{} and ranges from
\RVminSigI{} to \RVmaxI{}~bits, and a maximum-entropy bound estimated on
held-out sessions lies above it in all of them. Independent random
initialization reduces the leakage at 1024 shots from \RVmainI{} to
\RVindI{}~bits. For defense, we show that any
mechanism acting after the stopping decision can only add rounds, and
that differentially private padding costs the full adjacency radius. A
fixed public iteration budget removes the channel completely; set at the
median run length, it costs \RVbudFiftyPct{} of the remaining energy
gap. A differentially private stopping rule based on AboveThreshold
protects every pair of instances, but it must often stop too early,
which makes it costly.
\end{abstract}

\section{Introduction}
\label{sec:intro}
Blind quantum computation (BQC) lets a client with
limited quantum capability delegate a computation to a powerful,
untrusted server while concealing the input, the algorithm and the
output up to an explicitly declared leakage function. The construction
of Broadbent, Fitzsimons and Kashefi (BFK)~\cite{broadbent_universal_2009}
and its verifiable extension by Fitzsimons and Kashefi
(FK)~\cite{fitzsimons_unconditionally_2017} establish this for a
\emph{single} delegated computation, and Dunjko, Fitzsimons, Portmann
and Renner placed the guarantee on composable
footing~\cite{dunjko_composable_2014} within the abstract-cryptography
framework~\cite{maurer_abstract_2011,portmann_security_2022}.

Variational quantum algorithms (VQAs) are the setting in which much
near-term delegated quantum computing is expected to
occur~\cite{cerezo_variational_2021}. The variational quantum
eigensolver~\cite{peruzzo_variational_2014} and the quantum
approximate optimization algorithm~\cite{farhi_quantum_2014} delegate a
parametrized circuit, read out an estimated expectation value, and
update the parameters classically before delegating
again~\cite{mcclean_theory_2016}. So a single run delegates repeatedly, with a classical control process between rounds that
decides what to delegate next and, on the basis of private data, when
to stop.

Protocols for exactly this setting now exist. Inajetovic, Wallden and
Pappa give an end-to-end verifiable delegated VQA, proving that the optimization either returns a correct result or
aborts~\cite{inajetovic_verifiable_2026}; related constructions use
quantum homomorphic encryption~\cite{li_delegated_2023}. These works
inherit blindness from the underlying delegated protocol and add
convergence and soundness guarantees for the loop. They do not quantify
what the classical control flow of the loop reveals. Blindness concerns
the content of each round, whereas the number of rounds is a property
of the whole sequence.

\subsection{The channel}
A cloud quantum service knows how many jobs a client submitted in a
session and when each was submitted and returned. This is ordinary session and billing metadata. The provider obtains it
without any attack, and it lies outside the cryptographic boundary that
blindness protects. When the client runs a VQA, the number of jobs is a
public multiple of the number of optimizer iterations, and the number
of iterations is chosen by a stopping rule that reads the client's
private cost trajectory. The round count is therefore an output of the client's secret
computation, and the protocol reveals it openly.

Network security has studied the same situation for two decades under
the name traffic analysis. Hintz showed that the number, size and
timing of encrypted packets identify the website a user is visiting
without decrypting anything~\cite{hintz_fingerprinting_2003}, which started a large literature on website fingerprinting and on why
padding defenses against it often fail~\cite{dyer_peekaboo_2012}. We
ask whether the same kind of counting works against blind delegation,
how much it reveals in the variational setting, and which property of
the hidden problem it reflects.

This channel differs from the side channels studied so far in quantum
computing. Lu
\emph{et al.} recover structural information about a delegated circuit
from fine-grained execution timing on a commercial cloud quantum
service~\cite{lu_quantum_2025}, and power traces on quantum controllers
can reconstruct circuit
structure~\cite{xu_exploration_2023,erata_quantum_2024,bell_reconstructing_2022}.
Those attacks measure a physical signal during a single execution. The
channel studied here needs only the job log that a provider keeps
anyway, and it concerns the sequence of jobs, not the circuit.

\begin{table*}[!htbp]
\centering
\caption{Empirical hypotheses of Section~\ref{subsec:hypotheses}, the
test applied to each, and the outcome. Leakage is bias-corrected
plug-in mutual information in bits; all accuracies are measured on
held-out sessions.}
\label{tab:hypotheses}
\footnotesize
\begin{tabular}{@{}l>{\raggedright\arraybackslash}p{0.27\textwidth}>{\raggedright\arraybackslash}p{0.23\textwidth}>{\raggedright\arraybackslash}p{0.36\textwidth}@{}}
\toprule
& \textbf{Prediction} & \textbf{Test} & \textbf{Outcome} \\
\midrule
\ref{hyp:pl} & Decay factor $1-2\eta\mu_t$; early-window $\bar\mu$
forecasts $t^\star$ & Sec.~\ref{subsec:pl_results},
Table~\ref{tab:predictors}, Fig.~\ref{fig:pl} & Supported: decrement
error \RVdecTwo{} (\RVdecOne{} for $1-\eta\mu_t$); LOO error
\RVearlyLoo{} it.\ vs \RVnullLoo{} (constant), \RVgLoo{} ($g$),
\RVlamMinLoo{} ($\lambda^{+}_{\min}$). Ten instances, one family. \\
\ref{hyp:snr} & Held-out bound~\eqref{eq:ub} $\ge$ measured leakage &
Table~\ref{tab:systems}, Fig.~\ref{fig:universal} & Supported on
average in all \RVnCfgU{} configurations; per split in at least
\RVxfWorst{}. Measured/bound ratio \RVxfRatioLo{} to \RVxfRatioHi{}:
an envelope, not a predictor. \\
\ref{hyp:general} & Leakage survives changes of the control process
and the problem & Sec.~\ref{sec:generality}, Tables~\ref{tab:systems}
and~\ref{tab:rules} & Supported for optimizer, rule, initialization and
size. Partly refuted for the family: Max-Cut density leaks
\RVqgI{}~bits under gradient descent and nothing detectable under
SPSA; graph identity leaks nothing detectable. \\
\ref{hyp:transfer} & Mismatched adversary stays above chance &
Table~\ref{tab:transfer} & Supported across shot budgets
(\RVxsOffLo{} to \RVxsOffHi{}, chance $0.10$). Refuted across stopping
rules (mean \RVxrOffMean{}). \\
\ref{hyp:lecam} & Held-out error follows $\Phi(-\sqrt n\,\Delta t/2\sigma)$
& Fig.~\ref{fig:sessions}(b) & Partly supported: within
\RVtpMaxDiff{} percentage points for $n\le16$, but the Gaussian form
underestimates the tail. \\
\bottomrule
\end{tabular}
\end{table*}

\subsection{Formal results and empirical hypotheses}
\label{subsec:hypotheses}
We keep proved results separate from tested claims. The formal results
(Theorems~\ref{thm:tight}, \ref{thm:verif}, \ref{thm:ub},
\ref{thm:lb}, \ref{thm:dpi}, \ref{thm:impossible} and~\ref{thm:svt},
with the accompanying lemmas and propositions) hold under their stated
assumptions and need no experimental confirmation. The hypotheses below are empirical. Each makes a prediction that the
data could contradict, and Table~\ref{tab:hypotheses} reports the
outcomes, including the parts that were not supported.

\begin{hypothesis}[Landscape origin]
\label{hyp:pl}
The instance dependence of the stopping time is set by the deterministic
geometry of the cost landscape along the optimization path, summarized
by the local Polyak--{\L}ojasiewicz (PL) ratio $\mu_t$. Predictions:
(a) the per-step decay of the cost gap is $1-2\eta\mu_t$ to first order,
not the worst-case factor $1-\eta\mu_t$; (b) $\mu_t$ averaged over
iterations 5 to 15, which does not use the stopping round, forecasts
the noiseless stopping round out of sample better than a constant, the
Hessian spectrum at the optimum, or the secret parameter itself.
\end{hypothesis}

\begin{hypothesis}[Second-moment envelope]
\label{hyp:snr}
The maximum-entropy bound of Theorem~\ref{thm:ub}, computed from the
stopping-time means and variances of one half of the sessions, is at
least as large as the leakage measured on the other half, in every
configuration.
\end{hypothesis}

\begin{hypothesis}[Robustness]
\label{hyp:general}
The leakage remains significantly non-zero when the optimizer, the
stopping rule, the initialization, the system size and the problem
family are changed.
\end{hypothesis}

\begin{hypothesis}[Adversary transfer]
\label{hyp:transfer}
A likelihood adversary trained on a configuration that differs from the
client's retains above-chance identification accuracy.
\end{hypothesis}

\begin{hypothesis}[Two-point scaling]
\label{hyp:lecam}
The error of a held-out Bayes test between two instances from $n$
linked sessions follows the closed form of Theorem~\ref{thm:lb}.
\end{hypothesis}

\subsection{Contributions}
\begin{enumerate}
\item \textbf{A tight composition theorem} (Theorem~\ref{thm:tight}).
 Under fresh per-round randomness, the optimal advantage of an
 honest-but-curious server over an entire delegation sequence equals
 the total-variation distance between the stopping-time laws, and the
 server's information about any secret equals the information carried
 by the count alone (Corollary~\ref{cor:info}). The proof states the
one-time-pad step explicitly (Lemma~\ref{lem:pad}).
\item \textbf{Soundness under adaptive stopping}
 (Theorem~\ref{thm:verif}): an optional-stopping bound
 $\delta\,\Ex[T]$ in place of a union bound over a worst-case budget,
 \rev{under an explicit conditional soundness assumption
 (Assumption~\ref{ass:condsound}) that we show is met by composably
 secure verifiable protocols run with fresh secrets in every round}.
\item \textbf{Stopping-time laws and bounds.} The exact law for level-crossing rules; a proof that the two-window
slope rule used in practice is a level crossing at a threshold that
depends on the instance (Proposition~\ref{prop:slope}); a first-passage law for optimizer noise (Proposition~\ref{prop:ig});
a maximum-entropy bound under a Gaussian approximation, with a
counterexample showing that no bound based only on means and variances
holds in general (Theorem~\ref{thm:ub}, Remark~\ref{rem:counter}); a
two-point lower bound (Theorem~\ref{thm:lb}); and a monotonicity
theorem that needs no distributional assumption
(Theorem~\ref{thm:dpi}).
\item \textbf{Identification of the leaked quantity.} The round count
reflects the local PL ratio accumulated along the optimization path
(Proposition~\ref{prop:pl}). We test this with a forecast made early
in the run, which avoids circular reasoning
(Section~\ref{subsec:pl_results}).
\item \textbf{Measurements across \RVnCfgU{} configurations} with
 standard errors, held-out lower bounds, cross-fitted upper-bound tests,
 attacker-transfer matrices, an open-world evaluation over every
 monitored instance, and a decomposition of the stopping-time spread
 into readout and optimizer noise (Sections~\ref{sec:experiments}
 and~\ref{sec:generality}).
\item \textbf{Mitigations with guarantees and measured costs}
 (Section~\ref{sec:mitigation}): defenses that act after the stopping
decision can only pad
 (Lemma~\ref{lem:posthoc}) and differentially private padding is
 expensive for every $\eps$ (Theorem~\ref{thm:impossible}); a fixed
 public budget gives zero leakage (Proposition~\ref{prop:budget}); an
 implemented AboveThreshold stopping rule is $\eps$-differentially
 private at the instance level (Theorem~\ref{thm:svt}) but must stop
 prematurely with probability bounded below
 (Proposition~\ref{prop:premature}).
\end{enumerate}

\subsection{Scope and threat model}
\label{subsec:threat}
We work in the information-theoretic model of BFK- and FK-style
protocols. Our adversary is \emph{honest-but-curious}: the server
follows the protocol but records everything it can observe, including
metadata outside the cryptographic boundary. Its observation in a
session is
\begin{equation}
 \mathcal{O} \;=\; \big(\,\{\text{transcripts}\}_{t\le T},\, T \,\big),
 \label{eq:observation}
\end{equation}
and in the cross-session setting $\mathcal{O}^{(1)},\dots,\mathcal{O}^{(n)}$
for $n$ sessions it can link to one client. The client's secret is the
problem instance, abstracted as a parameter $\Theta$ with a prior known
to the server. Following Kerckhoffs' principle we let the server know
the client's software configuration (ansatz, optimizer, stopping rule,
shot budget); Section~\ref{subsec:transfer} measures what happens when
that knowledge is wrong. We do not model inter-round wall-clock timing,
which plausibly carries further leakage, and we do not measure on
commercial hardware; Section~\ref{sec:discussion} discusses both.

\section{Preliminaries}
\label{sec:prelim}

\subsection{Single-round blindness and verifiability}
Let $U$ be a target unitary, $\rho_{\mathrm{in}}$ the client's input,
and $\Pi$ a protocol terminating with the client accepting an output
or aborting.

\begin{definition}[Blindness]
\label{def:blindness}
$\Pi$ is \emph{blind up to $\Leak$} if, for every server strategy, the
server's reduced state at every point of the protocol depends on
$(U,\rho_{\mathrm{in}})$ only through $\Leak(U,\rho_{\mathrm{in}})$.
\end{definition}

\begin{definition}[Verifiability]
\label{def:verif}
$\Pi$ is \emph{$\delta$-verifiable} if, for every server deviation,
the probability that the client accepts an output further than a
negligible trace distance from $U\rho_{\mathrm{in}}U^{\dagger}$ is at
most $\delta$.
\end{definition}

BFK on an $n\times m$ brickwork state is blind in the sense of
Definition~\ref{def:blindness} with $\Leak(U,\rho_{\mathrm{in}})=(n,m)$. Each transmitted qubit is prepared
as $\ket{+_{\theta_j}}$ with $\theta_j$ uniform on the eight-element
set $\{k\pi/4\}$, and each measurement instruction is
$\delta_j=\phi'_j+\theta_j+\pi r_j$ with $r_j$ a uniform bit, where
$\phi'_j$ is the flow-corrected computation angle. The server returns
the outcome $s_j$, and the client decrypts $b_j=s_j\oplus r_j$.
Correctness holds for every value of the pads $(\theta_j,r_j)$.
Verifiability has been obtained in several ways. FK interleaves trap
qubits whose outcomes the client can predict and, with the computation
encoded in an error-correcting code of sufficient distance, makes the
probability of accepting a corrupted output exponentially small in the
code parameters~\cite{fitzsimons_unconditionally_2017}. For
classical-output (BQP) computations, Leichtle \emph{et al.} interleave
test rounds with repeated computation rounds and obtain a soundness
error exponentially small in the number of rounds, robust to a constant
level of device noise~\cite{leichtle_verifying_2021}; see
also~\cite{kapourniotis_unifying_2024}. We take such constructions as
given and write $\delta$ for the per-round soundness error of
Definition~\ref{def:verif}.

\subsection{Delegation sequences}
\begin{definition}[Delegation sequence]
\label{def:sequence}
A \emph{delegation sequence} is a triple $\mathcal{A}=(\{U_t\}_{t\ge1},C,T)$
in which each $U_t$ is delegated in round $t$ through an independent
instance of a protocol $\Pi$ blind up to $\Leak$; $C$ is a classical
control process producing $U_{t+1}$ from the decrypted outputs of
rounds $1,\dots,t$ and the client's private randomness; and $T$ is a
\emph{stopping time} with respect to the filtration
$(\mathcal{F}_t)_{t\ge0}$ generated by those outputs and that
randomness.
\end{definition}

Definition~\ref{def:sequence} covers a fixed circuit delegated a fixed number of times
($T$ constant) and a variational loop ($U_t=U(\theta_t)$ for a fixed
ansatz, $\theta_t$ updated by a classical optimizer that also decides
when to halt).

\begin{assumption}[Fresh randomness]
\label{ass:fresh}
In every round $t$, the pads $(\theta^{(t)}_j,r^{(t)}_j)_j$ are drawn
uniformly at random, independently of the pads of all other rounds and
of $\mathcal{F}_{t-1}$.
\end{assumption}

Assumption~\ref{ass:fresh} is an implementation requirement, analogous
to never reusing a stream-cipher keystream. A client that reuses pads
across rounds violates it, and nothing below applies.

\section{Composition and its exact limit}
\label{sec:composition}

\subsection{Blindness composes, and the bound is tight}
The key step is that, in each round, the server's record is
independent of what the client learns.
For an honest-but-curious server that follows BFK, every qubit it
receives is measured within the round, so its record of round $t$ once
that round has finished is the classical transcript
$V_t=(\delta^{(t)}_j,s^{(t)}_j)_j$. Let $B_t=(b^{(t)}_j)_j$ be the
client's decrypted outcomes in round $t$.

\begin{lemma}[One-time-pad factorization]
\label{lem:pad}
Under Assumption~\ref{ass:fresh}, for every $t$, conditionally on
$\mathcal{F}_{t-1}$ and on the server's records $V_1,\dots,V_{t-1}$,
the transcript $V_t$ is uniform on its alphabet and independent of
$(U_t,B_t)$. Consequently, for every $\tau$, the records
$(V_1,\dots,V_\tau)$ are independent of $\mathcal{F}_\tau$, with law
$\mu_1^{\otimes\tau}$ where $\mu_1$ is the uniform law determined by
$(n,m)$.
\end{lemma}

\begin{proof}
Fix the past. By correctness for every pad value, the conditional law
of $B_t$ given the pads is the computation's own outcome law, so $B_t$
is independent of the round-$t$ pads. Now condition on $B_t$ and $U_t$.
Then $s_j=b_j\oplus r_j$ is uniform because $r_j$ is, and for each value
of $r_j$ the angle $\delta_j=\phi'_j+\theta_j+\pi r_j$ is uniform on
the eight-element set because $\theta_j$ is. The pairs
$(\delta_j,s_j)$ are therefore uniform on an alphabet of size $16$ and,
since pads are independent across $j$, jointly uniform and independent
of $(U_t,B_t)$. The pads are fresh (Assumption~\ref{ass:fresh}), so
this holds conditionally on the whole past. Multiplying the
conditional laws over rounds gives the joint statement.
\end{proof}

\begin{theorem}[Tight sequence blindness]
\label{thm:tight}
Let $\mathcal{A}$ and $\mathcal{A}'$ be delegation sequences using the
same brickwork shape $(n,m)$ in every round and satisfying
Assumption~\ref{ass:fresh}, with stopping-time laws $P_T^{\mathcal A}$
and $P_T^{\mathcal A'}$. Let $\sigma^{\mathcal A}$ denote the law of
the honest-but-curious server's total record~\eqref{eq:observation}
under $\mathcal A$. Then
\begin{equation}
 \dTV\big(\sigma^{\mathcal A},\sigma^{\mathcal A'}\big)
 \;=\;
 \dTV\!\big(P_T^{\mathcal A},P_T^{\mathcal A'}\big).
 \label{eq:tight}
\end{equation}
\end{theorem}

\begin{proof}
\emph{Upper bound.} The event $\{T=\tau\}$ belongs to
$\mathcal{F}_\tau$, so by Lemma~\ref{lem:pad} the record factorizes as
$\sigma^{\mathcal A}(v_1,\dots,v_\tau,\tau)
=P^{\mathcal A}_T(\tau)\,\mu_1^{\otimes\tau}(v_1,\dots,v_\tau)$. Two
such laws differ only in the weights $P_T(\tau)$, and records with
different $\tau$ are disjoint events, so their total-variation distance
is that of the weights. This gives $\le$.

\emph{Lower bound.} The server reads $\tau$ from its record and applies
the Bayes-optimal test between $P_T^{\mathcal A}$ and
$P_T^{\mathcal A'}$, achieving advantage
$\dTV(P_T^{\mathcal A},P_T^{\mathcal A'})$. This gives $\ge$.
\end{proof}

Adaptivity could in principle break this argument, because $U_t$ is
chosen from earlier outcomes. It does not, because $U_t$ depends only
on the decrypted outcomes $B_{<t}$, and the pads make these independent of the server's records. Two
consequences follow directly: Corollary~\ref{cor:fixedT} treats a round
count fixed in advance, and Corollary~\ref{cor:info} restates
Theorem~\ref{thm:tight} in information-theoretic terms.

\begin{corollary}
\label{cor:fixedT}
If $T$ is a constant $\tau$ fixed in advance, the right-hand side
of~\eqref{eq:tight} vanishes and the sequence is blind up to
$\Leak=(n,m,\tau)$.
\end{corollary}

\begin{corollary}[Information form]
\label{cor:info}
For any secret $\Theta$ on which the client's computation depends, and
any prior, $\Iinfo(\mathcal{O};\Theta)=\Iinfo(T;\Theta)$, and for $n$
linked sessions run with independent pads,
$\Iinfo(\mathcal{O}^{(1)},\dots,\mathcal{O}^{(n)};\Theta)
=\Iinfo(T^{(1)},\dots,T^{(n)};\Theta)$.
\end{corollary}

\begin{proof}
By Lemma~\ref{lem:pad}, given $T$ the transcripts are distributed as
$\mu_1^{\otimes T}$ whatever $\Theta$ is, so
$\Theta\to T\to\mathcal{O}$ is a Markov chain; $T$ is also a function
of $\mathcal{O}$. The data-processing inequality in both directions
gives equality. The multi-session statement follows in the same way.
\end{proof}

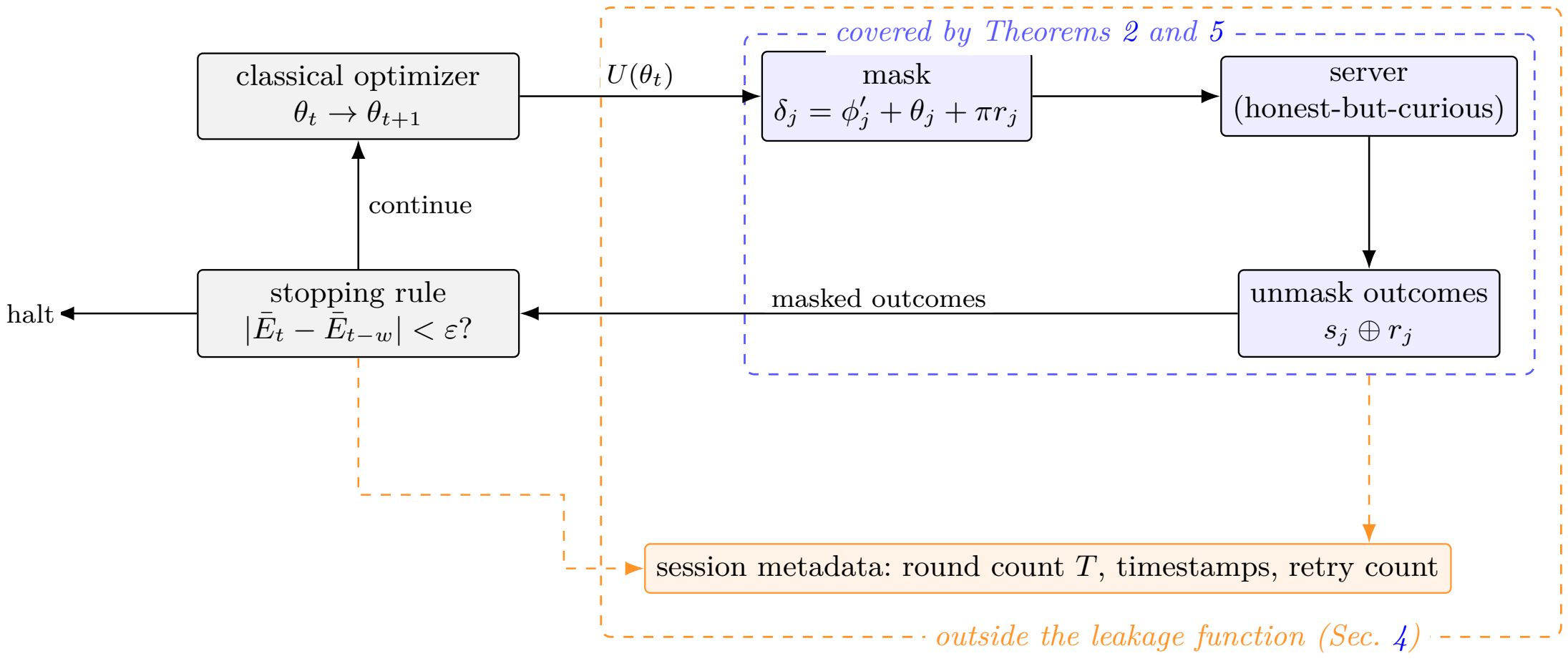
\begin{figure*}[!htbp]
\centering
\resizebox{0.95\textwidth}{!}{
\begin{tikzpicture}[
  x=1cm, y=1cm, font=\small,
  box/.style   = {draw, rounded corners=1.6pt, inner sep=3.4pt,
                  align=center, line width=0.5pt},
  client/.style= {box, fill=black!5},
  quant/.style = {box, fill=blue!7},
  meta/.style  = {box, fill=orange!10, draw=orange!80},
  arr/.style   = {-{Latex[length=2mm]}, line width=0.55pt},
  darr/.style  = {-{Latex[length=2mm]}, line width=0.55pt, dashed,
                  color=orange!85},
  elab/.style  = {font=\footnotesize, inner sep=1.8pt, fill=white},
  tag/.style   = {font=\small\itshape, inner sep=2.6pt, fill=white}]

\node[client, minimum width=3.4cm] (opt)  at (0,0)
  {classical optimizer\\$\theta_t \rightarrow \theta_{t+1}$};
\node[client, minimum width=3.4cm] (stop) at (0,-2.3)
  {stopping rule\\$|\bar E_t-\bar E_{t-w}| < \varepsilon$?};

\node[quant] (enc) at (5.7,0)
  {mask\\$\delta_j=\phi'_j+\theta_j+\pi r_j$};
\node[quant] (srv) at (10.7,0)
  {server\\(honest-but-curious)};
\node[quant] (dec) at (10.7,-2.3)
  {unmask outcomes\\$s_j \oplus r_j$};

\node[meta, minimum width=7.6cm] (mta) at (7.3,-5.0)
  {session metadata:\ round count $T$, timestamps, retry count};

\draw[arr] (opt) -- node[elab, above]{$U(\theta_t)$} (enc);
\draw[arr] (enc) -- (srv);
\draw[arr] (srv) -- (dec);
\draw[arr] (dec) -- node[elab, above]{masked outcomes} (stop);
\draw[arr] (stop) -- node[elab, right, xshift=1pt]{continue} (opt);
\draw[arr] (stop.west) -- ++(-1.5,0) node[elab, left, xshift=2pt]{halt};

\begin{scope}[on background layer]
  \node[draw=blue!65, dashed, line width=0.6pt, rounded corners=3pt,
        fit=(enc)(srv)(dec), inner sep=5pt] (inner) {};
  \node[draw=orange!85, dashed, line width=0.6pt, rounded corners=3pt,
        fit=(enc)(srv)(dec)(mta), inner sep=13pt] (outer) {};
\end{scope}

\draw[darr] (inner.south -| dec) -- (mta.north -| dec);
\draw[darr] (stop.south) -- ++(0,-1.45) -| ($(mta.west)+(-0.85,0)$)
            -- (mta.west);

\node[tag, text=blue!65, anchor=center]
  at ($(inner.north west)!0.36!(inner.north east)$)
  {covered by Theorems~\ref{thm:tight} and~\ref{thm:verif}};
\node[tag, text=orange!85, anchor=center]
  at ($(outer.south west)!0.60!(outer.south east)$)
  {outside the leakage function (Sec.~\ref{sec:channel})};
\end{tikzpicture}}
\caption{The two boundaries do not coincide. Everything inside the
inner boundary, meaning per-round angles, registers, outcome bits and
trap checks, is covered by Theorems~\ref{thm:tight}
and~\ref{thm:verif}. The round counter, the retry count and the stop or
continue decision sit outside it, in ordinary session metadata.
Theorem~\ref{thm:tight} says that the gap between the two boundaries is
exactly the stopping-time channel.}
\label{fig:boundary}
\end{figure*}

\begin{table}[!htbp]
\centering
\caption{What the composition theorems do and do not cover.}
\label{tab:summary}
\footnotesize
\begin{tabular}{@{}>{\raggedright\arraybackslash}p{0.40\textwidth}>{\raggedright\arraybackslash}p{0.44\textwidth}@{}}
\toprule
\textbf{Covered} (Thms.~\ref{thm:tight},~\ref{thm:verif}) &
\textbf{Not covered} (residual channel) \\
\midrule
Per-round angles $\delta^{(t)}_j$ & Round count $T$, when data-dependent \\
Per-round quantum registers & Inter-round wall-clock timing \\
Per-round outcome bits $s^{(t)}_j$ & Session-level scheduling metadata \\
Trap outcomes and accept or abort correctness & Anything correlated with $T$ across linked sessions \\
The public shape $(n,m)$ & Retry counts if Assumption~\ref{ass:abort} fails \\
\bottomrule
\end{tabular}
\end{table}

Corollary~\ref{cor:info} is the reason we measure
$\Iinfo(T;\Theta)$ in Sections~\ref{sec:experiments}
and~\ref{sec:generality}: in this model it equals the server's total information about the
secret. Figure~\ref{fig:boundary} illustrates the division that
Theorem~\ref{thm:tight} establishes between what blindness covers and
what it leaves exposed, and Table~\ref{tab:summary} lists the same
division item by item. Remark~\ref{rem:composable} explains how the
result relates to composable security.

\begin{remark}[Relation to composable security]
\label{rem:composable}
Theorem~\ref{thm:tight} is a stand-alone statement for an
honest-but-curious server. It does not contradict composable
security~\cite{dunjko_composable_2014}. In the abstract-cryptography
framework the ideal blind-computation resource leaks $(n,m)$ per call,
and a distinguisher interacting with the ideal system still sees how
many calls are made. The stopping-time channel is therefore present in
the ideal world as well, and composability certifies only that the real
protocol leaks nothing beyond it. If each real round is
$\eps_{\mathrm c}$-close to the ideal resource, a hybrid argument that
replaces round $t$ only on the event $\{T\ge t\}$ adds at most
$\eps_{\mathrm c}\sum_t\Pr[T\ge t]=\eps_{\mathrm c}\,\Ex[T]$ to the
left-hand side of~\eqref{eq:tight}. A malicious server can do more than
observe: by corrupting rounds it can influence $T$. Verification limits
undetected influence (Theorem~\ref{thm:verif}), but we do not analyze
that active channel here.
\end{remark}

\subsection{Verifiability under adaptive stopping}
Soundness needs separate treatment, because a server that deviates in
one adaptively chosen round out of many is a more capable adversary
than one restricted to a single round. Bounding the failure probability
by $\tau\delta$ for a fixed budget $\tau$ is valid but loose when the
round count is itself a stopping time.

\begin{revised}
Definition~\ref{def:verif} is a stand-alone statement: it bounds the
failure probability of one protocol instance run in isolation. An
adaptive server, however, enters round $s$ with a history: it has seen
the transcripts and the accept or abort decisions of earlier rounds,
and it may have deviated in them and kept classical or quantum memory.
A bound that holds for every round given that history does not follow
from Definition~\ref{def:verif} alone, so we state it as an
assumption. Let $\mathcal{H}_{s-1}$ be the $\sigma$-algebra generated
by everything that has happened before round $s$: the client's
decrypted outputs, its private randomness for rounds $1,\dots,s-1$, the
classical transcripts, the accept or abort decisions, and the classical
choices made by the server. It contains the filtration
$\mathcal{F}_{s-1}$ of Definition~\ref{def:sequence}, so $T$ is also a
stopping time with respect to $(\mathcal{H}_t)$.

\begin{assumption}[Conditional per-round soundness]
\label{ass:condsound}
For every server strategy and every round $s$, on the event
$\{T\ge s\}$,
\begin{equation}
 \Pr\big[F_s\mid\mathcal{H}_{s-1}\big]\;\le\;\delta
 \quad\text{almost surely,}
 \label{eq:condsound}
\end{equation}
where $F_s$ is the event that the client accepts an incorrect output in
round $s$.
\end{assumption}

Assumption~\ref{ass:condsound} holds for the protocols of
Section~\ref{sec:prelim} under two conditions, both of which are
standard. First, the client draws the secrets of round $s$ (the
one-time pads, and the positions and angles of the trap qubits or the
choice of test rounds) fresh and independently of $\mathcal{H}_{s-1}$,
which extends Assumption~\ref{ass:fresh} from the pads to the trap
randomness. Second, the stand-alone soundness bound holds against
every deviation of a server that starts the round holding arbitrary
side information, classical or quantum, that is independent of the
round's secrets. This second condition is what composable security
provides: in the abstract-cryptography framework the verifiable
protocol is $\delta$-close to an ideal resource that either returns the
correct output or aborts, for every distinguisher, including one that
has run earlier instances and kept their outputs and its own
memory~\cite{dunjko_composable_2014,maurer_abstract_2011,portmann_security_2022}.
For the verifiable protocols we rely on, this property is available as
follows. Dunjko \emph{et al.} reduce composable verifiability to local
criteria and sketch why the trap-based construction of Fitzsimons and
Kashefi meets
them~\cite{dunjko_composable_2014,fitzsimons_unconditionally_2017}, and
Leichtle \emph{et al.} show that their test-round construction is
composable and statistically secure~\cite{leichtle_verifying_2021}. Given both conditions,
fix any value of the history. The server then enters round $s$ with a
fixed strategy, chosen as a function of that history, and a memory
that is independent of the round's fresh secrets, which is exactly the
situation the stand-alone bound covers; this gives~\eqref{eq:condsound}.
A client that reuses trap positions or pads across rounds breaks the
first condition, and the assumption, and with it
Theorem~\ref{thm:verif}, can then fail.
\end{revised}

\begin{theorem}[Adaptive verifiability composition]
\label{thm:verif}
Let $\mathcal{A}$ be a delegation sequence in which every round runs an
independent instance of a $\delta$-verifiable protocol \rev{that
satisfies Assumption~\ref{ass:condsound}}. Fix any server strategy, including strategies that choose where to deviate adaptively,
and let $T$ be the resulting stopping time, with $\Ex[T]<\infty$ under
that strategy. Then
\begin{equation}
 \Pr\Big[\textstyle\bigcup_{t\le T}F_t\Big]\;\le\;\delta\,\Ex[T],
 \label{eq:wald}
\end{equation}
where $F_t$ is the event that the client accepts an incorrect output
in round $t$.
\end{theorem}

\begin{proof}
Let $N_t=\sum_{s\le t\wedge T}\mathbf{1}[F_s]$ and
$M_t=N_t-\delta\,(t\wedge T)$\rev{, both adapted to the filtration
$(\mathcal{H}_t)$. Assumption~\ref{ass:condsound} gives
$\Pr[F_s\mid\mathcal{H}_{s-1}]\le\delta$ on $\{T\ge s\}$ for every
server strategy, whatever happened before; the stand-alone
Definition~\ref{def:verif} alone would not suffice here}. Hence $M$ is a
supermartingale with respect to \rev{$(\mathcal{H}_t)$}
with increments bounded by one, and since $\Ex[T]<\infty$ the optional
stopping theorem gives $\Ex[N_T]\le\delta\,\Ex[T]$. Markov's inequality
yields $\Pr[N_T\ge1]\le\delta\,\Ex[T]$.
\end{proof}

Remark~\ref{rem:budget} spells out what Theorem~\ref{thm:verif} means
for provisioning soundness in practice.

\begin{remark}
\label{rem:budget}
Equation~\eqref{eq:wald} is stronger than a union bound over a
worst-case budget whenever the budget exceeds the mean run length. A
client that provisions $T_{\max}=\NUMTmaxEx$ iterations but typically
stops near $\Ex[T]\approx\NUMmeanTEx$ pays soundness proportional to the
latter. Because the expectation is taken under the server's strategy, a
client should cap $T$ at a public maximum, which makes
$\Ex[T]\le T_{\max}$ unconditionally. To hold an end-to-end budget
$\delta_{\mathrm{tot}}$, it suffices to provision per-round soundness
$\delta\le\delta_{\mathrm{tot}}/\Ex[T]$, which costs a number of extra
test rounds or traps logarithmic in $\Ex[T]$ for constructions whose
soundness error decays exponentially.
\end{remark}

\section{The stopping-time channel}
\label{sec:channel}

\subsection{Data-dependent stopping is the norm}
A variational loop halts on a rule such as ``stop when the change in
the estimated cost falls below $\eps$.'' Such a rule makes $T$ a
function of the observed cost sequence, hence of the parameter
trajectory, hence of the problem instance. Grouping jobs does not hide $T$ either: in our experiments each optimizer iteration requires
$2d+1=\NUMrounds$ delegated circuit evaluations for a
$d=\NUMdparams$-parameter ansatz under the parameter-shift rule, so the
billed job count is a known multiple of the iteration count and carries
identical information.

\subsection{Stopping-time laws}
We model the cost trajectory as converging geometrically,
\begin{equation}
 E_t-E_\star=\Delta_0\,\rho^{\,t}\,(1+\eta_t),
 \label{eq:geom}
\end{equation}
with $\rho\in(0,1)$ a local convergence rate, $\Delta_0$ the initial
gap and $\eta_t$ the relative readout error. Geometric convergence is
the regime of first-order methods under a PL
inequality~\cite{karimi_linear_2016}, it is the form in which
convergence of the delegated protocol of~\cite{inajetovic_verifiable_2026}
is proved, and finite-shot estimators for VQAs are unbiased with
variance proportional to
$1/N_{\mathrm{shots}}$~\cite{sweke_stochastic_2020}.
Section~\ref{sec:mechanism} identifies the landscape quantity that sets
$\rho$. For the level-crossing rule $T=\min\{t:E_t-E_\star<\eps\}$ the law of $T$ is available in closed form
(Proposition~\ref{prop:exactlaw}).

\begin{proposition}[Level-crossing law]
\label{prop:exactlaw}
If the $\eta_t$ are i.i.d.\ with c.d.f.\ $F$, then for every $t\ge1$
\begin{equation}
 \Pr[T>t]\;=\;\prod_{s=1}^{t}
 \Big[1-F\Big(\tfrac{\eps}{\Delta_0\rho^{\,s}}-1\Big)\Big].
 \label{eq:exactlaw}
\end{equation}
\end{proposition}

\begin{proof}
$\{T>t\}=\bigcap_{s\le t}\{\eta_s\ge \eps/(\Delta_0\rho^{s})-1\}$, and
the $\eta_s$ are independent.
\end{proof}

A level-crossing rule needs $E_\star$, which the client does not
know. Practical rules instead compare recent cost values with earlier
ones. The rule used in our experiments
halts at the first $t\ge2w$ with
\begin{equation}
 S_t=\frac1w\Big|\tfrac1w\!\!\sum_{s=t-w+1}^{t}\!\!\hat E_s
 -\tfrac1w\!\!\sum_{s=t-2w+1}^{t-w}\!\!\hat E_s\Big| < \eps .
 \label{eq:rule}
\end{equation}
Without readout noise it is an exact level crossing, at a threshold
that itself depends on the instance.

\begin{proposition}[Slope rule as level crossing]
\label{prop:slope}
Under~\eqref{eq:geom} with $\eta_t\equiv0$, for $t\ge2w$,
$S_t<\eps$ if and only if $E_t-E_\star<\eps_{\mathrm{eff}}(\rho)$, where
\begin{equation}
 \eps_{\mathrm{eff}}(\rho)=\eps\,\frac{w^2(1-\rho)\,\rho^{2w-1}}{(1-\rho^w)^2}
 \;\approx\;\frac{\eps}{1-\rho}\quad (w(1-\rho)\ll1).
 \label{eq:epseff}
\end{equation}
\end{proposition}

\begin{proof}
Summing the geometric series, the two window means are
$\Delta_0\rho^{t-w+1}(1-\rho^w)/(w(1-\rho))$ and
$\Delta_0\rho^{t-2w+1}(1-\rho^w)/(w(1-\rho))$, so
$S_t=\Delta_0\rho^{t-2w+1}(1-\rho^w)^2/(w^2(1-\rho))$, which is below
$\eps$ exactly when $\Delta_0\rho^t<\eps_{\mathrm{eff}}(\rho)$. The
approximation uses $1-\rho^w\approx w(1-\rho)$ and $\rho^{2w-1}\approx1$.
\end{proof}

With $\eps_{\mathrm{eff}}$ given by~\eqref{eq:epseff}, the noiseless
stopping round is
\begin{equation}
 t^\star=\frac{\log\big(\Delta_0/\eps_{\mathrm{eff}}(\rho)\big)}{\log(1/\rho)} .
 \label{eq:tstar}
\end{equation}
In~\eqref{eq:tstar}, slow convergence acts in two directions. The denominator
shrinks, which lengthens the run, and the effective threshold rises,
which shortens it. The second effect is only logarithmic, so slower
instances still stop later. With readout
noise, $S_t$ is a difference of window averages, so the noise enters with reduced variance and
law~\eqref{eq:exactlaw} does not apply directly; we use it only for the
level-crossing rule.

Readout noise is not the only source of spread. The optimizer's own
noise (shot noise in gradient estimates, and the spread of initial
points) makes the trajectory itself random, and its effect on $T$ has a
different shape.

\begin{proposition}[Optimizer noise gives right skew]
\label{prop:ig}
Suppose the log-gap $X_t=\log(E_t-E_\star)$ follows the diffusion
approximation $dX=-a\,dt+s\,dW$ with $a>0$, and the rule fires at first
passage to $x_{\mathrm{eff}}=\log\eps_{\mathrm{eff}}$ from
$x_0=\log\Delta_0$, with $b=x_0-x_{\mathrm{eff}}>0$. Then $T$ is inverse
Gaussian with mean $b/a$, variance $bs^2/a^3$ and skewness
$3s/\sqrt{ab}>0$.
\end{proposition}

\begin{proof}
The first-passage time of Brownian motion with drift $-a$ and diffusion
$s$ to a level at distance $b$ is inverse Gaussian with mean $b/a$ and
shape $b^2/s^2$~\cite{folks_inverse_1978}. Its variance is
mean$^3$/shape and its skewness is $3\sqrt{\text{mean}/\text{shape}}$.
\end{proof}

The two noise sources predict opposite skews. Under
law~\eqref{eq:exactlaw}, multiplicative readout noise can make the cost
cross the threshold early by chance, but rarely much later than the
noiseless crossing. For representative
parameters\footnote{$\Delta_0=3$, $\rho=0.93$, $\eta_t\sim\mathcal N(0,0.08^2)$,
$\eps=0.05$; $6\times10^4$ Monte Carlo runs.} this law has skewness
$\RVskewLC$. Optimizer noise gives the positive skew of
Proposition~\ref{prop:ig}. Section~\ref{subsec:noise} uses this difference to identify the source
of the measured spread.

For closed-form statements we also use the Gaussian approximation
\begin{equation}
 T\mid\Theta \;\approx\; \mathcal{N}\!\big(t^\star(\Theta),\,
 \sigma_\Theta^2\big),
 \label{eq:surrogate}
\end{equation}
and test it in Section~\ref{subsec:gof}.

\subsection{An upper bound under the Gaussian approximation}
A tempting shortcut is to insert a Fisher information into the
capacity formula of a Gaussian channel. That step is not valid here,
because the noise variance depends on the parameter. The correct
statement under the Gaussian approximation is the following.

\begin{theorem}[Maximum-entropy bound]
\label{thm:ub}
Let $\Theta$ have any prior and let $T$ satisfy~\eqref{eq:surrogate}
exactly. Then
\begin{equation}
 \Iinfo(T;\Theta)\;\le\;
 \tfrac12\log_2\kappa,\qquad
 \kappa=\frac{\Var_\Theta\!\big[t^\star(\Theta)\big]+\Ex_\Theta[\sigma^2_\Theta]}
 {\bar\sigma^2},
 \label{eq:ub}
\end{equation}
where $\bar\sigma^2 = \exp\!\big(\Ex_\Theta[\ln\sigma^2_\Theta]\big)$
is the geometric mean of the conditional variances. We write
$\SNR_{\mathrm{eff}}=\sqrt{\kappa-1}$, so that~\eqref{eq:ub} reads
$\frac12\log_2(1+\SNR_{\mathrm{eff}}^2)$; if $\sigma_\Theta\equiv\sigma$,
$\SNR_{\mathrm{eff}}^2=\Var_\Theta[t^\star]/\sigma^2$.
\end{theorem}

\begin{proof}
We have $\Iinfo(T;\Theta)=h(T)-h(T\mid\Theta)$. The Gaussian maximizes
differential entropy at fixed
variance~\cite[Thm.~8.6.5]{cover_elements_2006}, and
under~\eqref{eq:surrogate} the conditional entropy is exact, so
\begin{gather*}
 h(T)\le\tfrac12\log_2(2\pi e\Var T),\\
 h(T\mid\Theta)=\Ex_\Theta\big[\tfrac12\log_2(2\pi e\sigma^2_\Theta)\big].
\end{gather*}
Subtracting gives
$\frac12\log_2(\Var T/\bar\sigma^2)$, and the law of total variance
gives $\Var T=\Var_\Theta[t^\star]+\Ex_\Theta[\sigma^2_\Theta]$.
\end{proof}

\begin{remark}[No two-moment bound holds in general]
\label{rem:counter}
Theorem~\ref{thm:ub} needs the Gaussian approximation, and no bound that depends
only on the conditional means and variances can replace it. Let
$T\mid\{\Theta=0\}$ be uniform on $\{1,3\}$ and let $T\mid\{\Theta=1\}$
put mass $\frac18,\frac34,\frac18$ on $\{0,2,4\}$. Both laws have mean
$2$ and variance $1$, so $\kappa=1$ and~\eqref{eq:ub} gives zero, yet the
supports are disjoint and $\Iinfo(T;\Theta)=1$~bit. One can always
replace $H(T)$ by its maximum-entropy bound
$\frac12\log_2(2\pi e(\Var T+\frac1{12}))$ for integer $T$, but the
resulting inequality still requires the conditional entropies
$H(T\mid\Theta)$, which are as hard to estimate as the mutual
information itself. We therefore treat~\eqref{eq:ub}, evaluated on
sample moments, as a prediction to be tested (Hypothesis~\ref{hyp:snr})
and not as a certificate.
\end{remark}

\subsection{A matching operational lower bound}
How well a server can actually do has an exact answer when there are
only two candidate instances.

\begin{theorem}[Two-point identifiability]
\label{thm:lb}
Let the client's instance be one of two possibilities with equal prior,
inducing stopping-time laws $P_0,P_1$, and let the server observe $n$
independent sessions. The minimum error probability of any test is
$P_e^{(n)}=\frac12\big(1-\dTV(P_0^{\otimes n},P_1^{\otimes n})\big)$,
and $\Iinfo(T^{n};\Theta)\ge 1-h_2\big(P_e^{(n)}\big)$ bits, where
$h_2$ is the binary entropy. Under~\eqref{eq:surrogate} with common
$\sigma$ and separation $\Delta t=|t^\star_1-t^\star_0|$,
\begin{equation}
 P_e^{(n)} \;=\; \Phi\!\Big(-\tfrac{\sqrt{n}\,\Delta t}{2\sigma}\Big),
 \label{eq:lecam}
\end{equation}
so the error is at most $\beta$ once
\begin{equation}
 n \;\ge\; \Big(\tfrac{2\sigma\,\Phi^{-1}(\beta)}{\Delta t}\Big)^{2}.
 \label{eq:nsessions}
\end{equation}
\end{theorem}

\begin{proof}
The first equality is the Le Cam characterization of Bayes risk for a
uniform binary prior~\cite[Ch.~2]{tsybakov_introduction_2009}. For two
Gaussians with equal variance, the sample mean is sufficient and
distributed as $\mathcal{N}(t^\star_i,\sigma^2/n)$, and the optimal test
thresholds at the midpoint, giving~\eqref{eq:lecam}. Inverting
gives~\eqref{eq:nsessions}. The information statement is Fano's
inequality: $H(\Theta\mid T^n)\le h_2(P_e^{(n)})$ and $H(\Theta)=1$.
\end{proof}

Mutual information saturates at $\log_2$ of the number of hypotheses,
while~\eqref{eq:lecam} falls exponentially in $n$. A channel worth a
fraction of a bit per session can still identify
well-separated instances reliably after a few tens of sessions, a
number of sessions that a cloud provider can easily observe.

\subsection{Retries, jitter, and a distribution-free monotonicity}
\label{subsec:retries}
A verifiable protocol aborts some rounds, more often because device
noise disturbed a test than because the server cheated, and a client
that cannot tell the difference retries. Schedulers also insert
queueing delays.

\begin{assumption}[Instance-independent abort rate]
\label{ass:abort}
The probability that a round aborts from device noise depends on the
hardware and the brickwork shape $(n,m)$ but not on the algorithmic
angles delegated in that round.
\end{assumption}

\begin{theorem}[Monotonicity under independent perturbation]
\label{thm:dpi}
Let $T_{\mathrm{obs}}=f(T,R)$ where $R$ is any random variable
independent of $\Theta$ given $T$ and $f$ is measurable, for example
$T_{\mathrm{obs}}=T+R$ for a retry count $R$ satisfying
Assumption~\ref{ass:abort} or for deliberate padding. Then
\begin{equation}
 \Iinfo(T_{\mathrm{obs}};\Theta)\le\Iinfo(T;\Theta),
 \quad
 \dTV\big(P_0^{\mathrm{obs}},P_1^{\mathrm{obs}}\big)\le\dTV(P_0,P_1).
 \label{eq:dpi}
\end{equation}
\end{theorem}

\begin{proof}
$\Theta\to T\to T_{\mathrm{obs}}$ is a Markov chain, and both
inequalities are the data-processing inequality~\cite{cover_elements_2006}.
\end{proof}

Inequality~\eqref{eq:dpi} needs no distributional assumption. Under
Assumption~\ref{ass:abort}, retries only dilute the existing channel and
do not open a second one, and the theorem also covers every mitigation
in Section~\ref{sec:mitigation} that perturbs a count after it has been
fixed. On hardware whose error rates depend on the delegated angles,
however, Assumption~\ref{ass:abort} fails and retry counts can become a
second channel.

\section{What the channel carries}
\label{sec:mechanism}

Sections~\ref{sec:composition} and~\ref{sec:channel} show that the
round count is the residual channel and describe its distribution. This
section asks which property of the cost landscape the count reflects.

\subsection{The local Polyak--{\L}ojasiewicz ratio}
For a cost function $E$ with minimum value $E_\star$ on the reachable
set, define along a trajectory $(\theta_t)$ the local
Polyak--{\L}ojasiewicz ratio
\begin{equation}
 \mu_t \;=\;
 \frac{\|\nabla E(\theta_t)\|^2}{2\big(E(\theta_t)-E_\star\big)} .
 \label{eq:pl}
\end{equation}
The PL inequality $\|\nabla E\|^2\ge 2\mu(E-E_\star)$ with a global
constant $\mu=\inf_t\mu_t$ was introduced by
Polyak~\cite{polyak_gradient_1963} and shown by Karimi, Nutini and
Schmidt to be weaker than the other conditions commonly used to prove
linear convergence while still sufficient for
it~\cite{karimi_linear_2016}. The ratio~\eqref{eq:pl} is its local,
trajectory-dependent version. It needs no convexity and no assumption
on the rank of the Hessian.

\begin{proposition}[PL accounting]
\label{prop:pl}
Let $\theta_{t+1}=\theta_t-\eta\nabla E(\theta_t)$ with $E$ being
$L$-smooth. Then
\begin{equation}
 \frac{E_{t+1}-E_\star}{E_t-E_\star}=1-2\eta\mu_t+r_t,
 \qquad |r_t|\le\eta^2L\mu_t .
 \label{eq:placc}
\end{equation}
If $\eta\le1/L$, the ratio is at most $1-\eta\mu_t$.
\end{proposition}

\begin{proof}
By Taylor's theorem with Lagrange remainder,
\[
 E_{t+1}=E_t-\eta\|\nabla E_t\|^2
 +\tfrac{\eta^2}{2}\nabla E_t^{\top}H(\xi)\nabla E_t ,
\]
with $|\nabla E_t^{\top}H(\xi)\nabla E_t|\le L\|\nabla E_t\|^2$.
Dividing by $E_t-E_\star$ and using~\eqref{eq:pl} gives the identity
and the bound on $r_t$. The upper bound
$1-2\eta\mu_t+\eta^2L\mu_t=1-\eta\mu_t(2-\eta L)\le1-\eta\mu_t$ for
$\eta L\le1$ is the descent-lemma guarantee.
\end{proof}

The two factors differ by a factor of two in the decrement. The
familiar $1-\eta\mu$ is a worst-case guarantee, while a real trajectory
with small $\eta L$ follows the first-order value $1-2\eta\mu_t$.
Summing the logarithm of~\eqref{eq:placc},
\begin{equation}
 \log\frac{E_t-E_\star}{\Delta_0}\;\approx\;-2\eta\sum_{s<t}\mu_s ,
 \label{eq:plsum}
\end{equation}
and combining~\eqref{eq:plsum} with Proposition~\ref{prop:slope} and
\eqref{eq:tstar}, the noiseless stopping round is the first $t$ at which $2\eta\sum_{s<t}\mu_s$ reaches
$\log(\Delta_0/\eps_{\mathrm{eff}})$. For a slowly varying ratio,
$t^\star\approx\log(\Delta_0/\eps_{\mathrm{eff}})/(2\eta\bar\mu)$. In short, the round count reflects the accumulated PL ratio along the
path, together with the initial gap. Remark~\ref{rem:barren} connects this
quantity to trainability.

\begin{remark}
\label{rem:barren}
The PL ratio is tied to trainability. In the barren-plateau regime the
gradient is exponentially suppressed in system size at fixed cost
gap~\cite{larocca_barren_2025}, and by~\eqref{eq:pl} $\mu_t$ is
suppressed with it. Problems that are hard to train produce long sessions, and the
provider can see this without decrypting anything.
\end{remark}

\subsection{Why the Hessian at the optimum is the wrong predictor}
A reader familiar with quadratic optimization might expect the rate to
be set by the smallest non-zero eigenvalue $\lambda^{+}_{\min}$ of the
Hessian $H$ at the optimum, since on a quadratic the gap component
along eigendirection $i$ contracts by $(1-\eta\lambda_i)^2$ per step and
directions with $\lambda_i=0$ carry no gap. That argument describes the asymptotic regime, and the stopping rule
fires long before it is reached. We computed $H$ exactly at the converged parameters for each
instance of Section~\ref{sec:experiments}: with $d=\NUMhessD$
parameters, between \RVrankLo{} and \RVrankHi{} eigenvalues exceed
$10^{-3}$, and the smallest eigenvalue above $10^{-2}$ lies between
\RVlamMinLo{} and \RVlamMinHi{}. The corresponding contraction time
$1/(2\eta\lambda^{+}_{\min})$ is \RVtauSlowLo{} to \RVtauSlowHi{}
iterations, an order of magnitude longer than any run. Before the stop, the gap decays mainly along stiffer directions, so
the gap-weighted quantity~\eqref{eq:pl} is the better summary.
The forecasting test of Section~\ref{subsec:pl_results} confirms that $\lambda^{+}_{\min}$ predicts
$t^\star$ worse than a constant.

\section{Experimental evaluation}
\label{sec:experiments}

\subsection{Setup}
We simulate a delegated variational eigensolver end to end. The client
seeks the ground state of the one-dimensional transverse-field Ising
model with open boundaries on $n=\NUMnqubits$ sites,
\begin{equation}
 H(g) \;=\; -\sum_{i=1}^{n-1} Z_iZ_{i+1}\;-\;g\sum_{i=1}^{n} X_i ,
 \label{eq:tfim}
\end{equation}
The client's secret is the transverse field $g$ in~\eqref{eq:tfim},
which takes $10$ equiprobable
values $g\in\{0.2,0.4,\dots,2.0\}$. The model spans an ordered phase, a
finite-size remnant of the critical point and a disordered
phase~\cite{sachdev_quantum_2011}, so the difficulty of the
optimization varies smoothly across the instance set.

The client runs a hardware-efficient ansatz with $L=\NUMnlayers$ layers
of $R_Y$ and $R_Z$ rotations and a linear CNOT chain,
$d=\NUMdparams$ parameters, and performs gradient descent with learning
rate $\eta=\NUMlr$ using parameter-shift gradients. $H(g)$ splits into
two commuting groups, measured in the computational and the Hadamard
basis, and each group estimator is the sample mean of
$N_{\mathrm{shots}}$ readouts. $N_{\mathrm{shots}}$ is the sampling
budget per expectation value; it is not a model of device noise, which
we do not simulate. Every iteration consumes $2d+1=\NUMrounds$ delegated
rounds. The stopping rule is~\eqref{eq:rule} with $w=\NUMwindow$ and
$\eps=\NUMepsstop$, and the budget is $T_{\max}=\NUMmaxiter$ iterations.
No session in any configuration reached its budget.

In the main configuration every session starts from a common parameter
vector, drawn once from $\mathcal N(0,0.35^2)$ per component, plus
per-session jitter $\mathcal N(0,0.05^2)$. This models a client that reuses a fixed initialization, which is
common in practice and favorable to the adversary. Section~\ref{sec:generality} repeats the experiment
with independent initializations. We run $\NUMrunsmain$ sessions per
instance at $N_{\mathrm{shots}}=1024$ and $\NUMrunsother$ per instance
at each of $256$, $4096$ and $16384$ shots.

\subsection{Statistical methodology}
\label{subsec:stats}
\emph{Leakage.} We report the plug-in mutual information between $T$
and the secret minus the mean of $200$ label-permutation replicates.
The permutation mean estimates the finite-sample bias under
independence; it is a heuristic correction, not an unbiased estimator
under dependence, and we therefore also report a rigorous lower bound
(below). The permutation $p$-value is $(1+\#\{\text{null}\ge I\})/201$,
so the smallest attainable value is $0.005$.

\emph{Uncertainty.} Standard errors come from a stratified bootstrap
that resamples sessions within each instance and recomputes the
permutation correction on every replicate; we quote $I\pm1.96\,$SE.

\emph{Held-out lower bound.} For any conditional model
$q(\theta\mid t)$,
$\Iinfo(T;\Theta)\ge H(\Theta)+\Ex[\log_2 q(\Theta\mid T)]$
(Barber and Agakov~\cite{barber_information_2003}). We take $q$ to be the Bayes posterior built from class-conditional
histograms (with half a count added to every bin) on one half of the
sessions and evaluate the expectation on the other
half, over $20$ random stratified splits. The estimate carries no
overfitting bias and is a valid lower bound in expectation.

\emph{Accuracy.} All adversary accuracies are measured on sessions not
used to fit the adversary's likelihoods.

\emph{Cross-fitting.} To test Hypothesis~\ref{hyp:snr} without using
the same data twice, we split sessions into halves A and B, evaluate
bound~\eqref{eq:ub} from the sample moments of A and the
bias-corrected leakage on B, and repeat over $40$ stratified splits.

\emph{Gaussianity.} For each instance we test the Gaussian
approximation~\eqref{eq:surrogate} with a Kolmogorov--Smirnov test in
which the mean and variance are
estimated from the data, as in the Lilliefors
test~\cite{lilliefors_kolmogorov_1967}. To remove ties, we first add
independent noise uniform on $[-\frac12,\frac12]$ to the counts, and
we obtain $p$-values from $400$ parametric Monte Carlo samples.

\begin{figure*}[!htbp]
\centering
\begin{minipage}[b]{0.485\textwidth}\centering
\includegraphics[width=\linewidth]{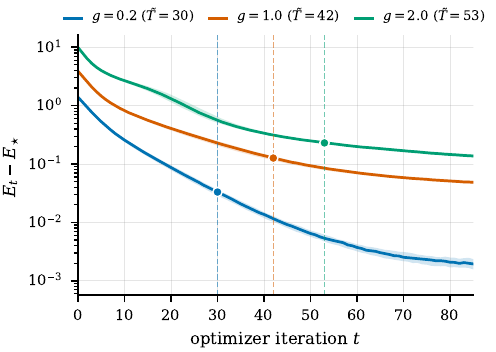}\\[-1pt]
{\footnotesize (a)}
\end{minipage}\hfill
\begin{minipage}[b]{0.485\textwidth}\centering
\includegraphics[width=\linewidth]{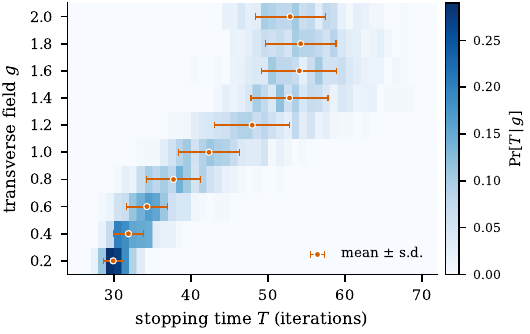}\\[-1pt]
{\footnotesize (b)}
\end{minipage}
\caption{Stopping times of the delegated eigensolver at $1024$ shots.
(a) Cost trajectories for three instances, shown as the gap to the
reachable variational optimum on a logarithmic scale. Shaded bands are
the interquartile range over sessions; markers and dashed lines give
the median stopping times. The per-round transcripts are identically
distributed in all three cases (Lemma~\ref{lem:pad}); only the round at
which the client stops differs. (b) Conditional stopping-time law
$\Pr[T\mid g]$ over sessions (heat map), with the mean and one standard
deviation overlaid.}
\label{fig:traj}
\end{figure*}

\subsection{The stopping time tracks the landscape}
\label{subsec:pl_results}
Figure~\ref{fig:traj}(a) shows the cost trajectories. The optimization
slows as $g$ increases toward and beyond the finite-size critical
region, and the rule fires correspondingly later: mean stopping times
run from $\NUMTlo$ iterations at $g=0.2$ to $\NUMThi$ at $g=\NUMghi$,
a separation of $\NUMTspread$ iterations or $\NUMroundspread$ delegated
rounds. The conditional laws $\Pr[T\mid g]$ (Fig.~\ref{fig:traj}(b)) are
distinct but overlapping. The mean measured stopping time agrees with
the stopping round of the noiseless trajectory to within
\NUMnlDev{}~iterations across the family (slope \NUMnlSlope{},
$R^2=\NUMnlRsq$), so readout noise sets the width of $\Pr[T\mid g]$ and
the landscape its location.

We test Hypothesis~\ref{hyp:pl} on noiseless gradient-descent
trajectories from the common initial point, one per instance, on which
$\mu_t$ can be computed exactly with $E_\star$ the reachable
variational minimum. Prediction (a) concerns the decay of the gap. Over
all iterations before the stop, $1-2\eta\mu_t$ reproduces the measured
per-step decrement $1-(E_{t+1}-E_\star)/(E_t-E_\star)$ with mean
relative error \RVdecTwo{}, whereas the worst-case factor
$1-\eta\mu_t$ misses it by \RVdecOne{}, the factor of two of
Proposition~\ref{prop:pl}. Because the per-step ratio is close to one,
an error measured on the ratio itself would hide this: on that scale
$1-\eta\mu_t$ is off by only \RVdecOneRatio{}.

\begin{figure}[!htbp]
\centering
\includegraphics[width=0.62\textwidth]{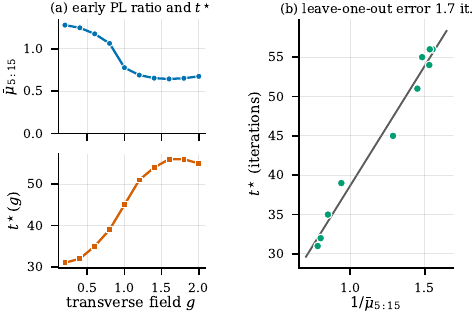}
\caption{An early-window forecast of the stopping round
(Hypothesis~\ref{hyp:pl}). (a) The local PL ratio averaged over
iterations 5 to 15, and the noiseless stopping round $t^\star$, against
the transverse field. (b) $t^\star$ against the reciprocal of the early
ratio, with the affine fit; the title gives the leave-one-out error.
The intercept absorbs the instance-independent part of the initial
approach.}
\label{fig:pl}
\end{figure}

\begin{table}[!htbp]
\centering
\caption{Forecasting the noiseless stopping round $t^\star$ from ten
instances by an affine fit in each predictor. LOO RMSE is the
leave-one-out root-mean-square error in iterations. Hessian quantities
are at the converged parameters; $\lambda^{+}_{\min}$ is the smallest
eigenvalue above $10^{-2}$.}
\label{tab:predictors}
\footnotesize
\begin{tabular}{@{}lccc@{}}
\toprule
Predictor of $t^\star$ & Pearson & Spearman & LOO RMSE \\
\midrule
$1/\bar\mu$, iterations 5 to 15 & $+0.990$ & $+0.985$ & 1.67 \\
$1/\|\nabla E\|^2$, iterations 5 to 15 & $-0.961$ & $-0.960$ & 3.90 \\
initial gap $\Delta_0$ & $+0.934$ & $+0.960$ & 4.72 \\
$\lambda_{\max}$ of $H$ at optimum & $+0.819$ & $+0.960$ & 7.62 \\
$1/\lambda_{\min}^{+}$ of $H$ at optimum & $+0.310$ & $+0.298$ & 11.43 \\
$1/\bar\mu$, stopping window & $+0.948$ & $+0.815$ & 3.78 \\
the secret $g$ itself & $+0.960$ & $+0.960$ & 3.57 \\
constant (no predictor) & n/a & n/a & 10.32 \\
\bottomrule
\end{tabular}

\end{table}

Prediction (b) is tested as a forecast. A correlation between
$t^\star$ and $\mu$ averaged over the iterations just before the stop
would be partly circular, because the stopping rule fires when progress
slows, and $\mu_t$ falls at the same time. We instead average $\mu_t$ over iterations 5 to 15,
long before any run stops ($t^\star\ge\NUMTlo$), and forecast
$t^\star$ by an affine fit scored by leave-one-out (LOO) error, since
with ten instances a two-parameter $R^2$ says little.
Table~\ref{tab:predictors} and Fig.~\ref{fig:pl} give the result. The
early-window ratio, which varies by a factor of about \RVmuEarlyRatio{}
across the family, forecasts $t^\star$ with LOO error
\RVearlyLoo{}~iterations (Pearson \RVearlyR{}, Spearman \RVearlyRho{}),
against \RVnullLoo{} for a constant, \RVgLoo{} for the secret $g$
itself, \RVstopLoo{} for the stopping-window ratio, and
\RVlamMinLoo{} for $1/\lambda^{+}_{\min}$ (\RVlamMinLooB{} with a
$10^{-3}$ eigenvalue cut). Within the disordered phase ($g\ge1.2$),
where $t^\star$ varies least, the early-window ratio keeps the correct
rank order (Spearman $\RVwithinEarly$), while the stopping-window ratio reverses
it ($\RVwithinStop$). The gradient norm alone predicts worse
than the ratio, because larger initial gaps come with larger gradients.
The test uses ten instances of one family and one trajectory per
instance. It shows that, in this family, the early landscape geometry
largely determines when the client stops. It does not show that the
same relation holds in other families.

\begin{table}[!htbp]
\centering
\caption{Leakage of the round count about the transverse field over
$10$ equiprobable instances ($\log_2 10=3.32$ bits). Measured values
are bias-corrected plug-in estimates $\pm1.96$ bootstrap SE; the
held-out lower bound is that of Section~\ref{subsec:stats}; the upper
bound is~\eqref{eq:ub} evaluated on full-sample moments. All permutation $p\le0.005$.}
\label{tab:leakage}
\footnotesize
\begin{tabular}{@{}rccc@{}}
\toprule
Shots & $I(T;g)$, 95\% & Held-out LB & Upper bound \\
\midrule
256 & $0.830\pm0.058$ & 0.614 & 1.104 \\
1024 & $1.236\pm0.038$ & 1.097 & 1.545 \\
4096 & $1.463\pm0.050$ & 1.270 & 1.976 \\
16384 & $1.684\pm0.058$ & 1.503 & 2.265 \\
\bottomrule
\end{tabular}

\end{table}

\begin{figure}[!htbp]
\centering
\includegraphics[width=0.62\textwidth]{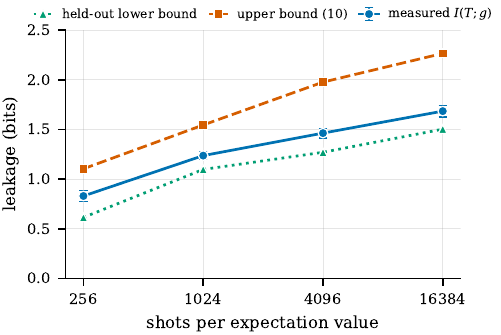}
\caption{Leakage versus shot budget, with $\pm1.96$ SE error bars, the
held-out lower bound, and bound~\eqref{eq:ub}. Larger shot budgets
narrow the stopping-time laws and widen the channel.}
\label{fig:leakage}
\end{figure}

\subsection{Measured leakage and the bounds}
Table~\ref{tab:leakage} and Fig.~\ref{fig:leakage} report the leakage
of the round count about $g$. At $1024$ shots the count carries
$\RVmainI\pm\RVmainSE$~bits (one standard error), with a held-out lower
bound of \RVmainBA{}~bits, and a held-out adversary identifies $g$
exactly in \RVaccMain{} of single sessions against a chance level of
$10\%$, and to within one grid step in \NUMaccTolOne{}. Leakage grows
with the shot budget, from $0.83$~bits at $256$ shots to
$\RVmaxI$~bits at $16384$, because the conditional laws narrow while
their locations barely move.
Bound~\eqref{eq:ub}, evaluated on the full-sample moments, lies above
every measurement.

\begin{table}[!htbp]
\centering
\caption{Decomposition of the within-class spread of $T$. ``Process''
replays the stopping rule on exact energies along the same trajectories,
removing readout noise from the stopping decision only. Share is the
process fraction of the within-class variance. Skewness is averaged
over instances.}
\label{tab:noise}
\footnotesize
\begin{tabular}{@{}rccccc@{}}
\toprule
& \multicolumn{2}{c}{Within-class s.d.} & Process & \multicolumn{2}{c}{Skewness} \\
\cmidrule(lr){2-3}\cmidrule(l){5-6}
Shots & full & process & share & full & process \\
\midrule
256 & 5.09 & 2.71 & 28\% & $+0.33$ & $+0.50$ \\
1024 & 3.94 & 2.37 & 36\% & $+0.35$ & $+0.43$ \\
4096 & 3.15 & 2.23 & 50\% & $+0.17$ & $+0.29$ \\
16384 & 2.56 & 2.27 & 79\% & $+0.40$ & $+0.57$ \\
\bottomrule
\end{tabular}

\end{table}

\subsection{Where the spread comes from}
\label{subsec:noise}
To separate readout noise from optimizer noise, we replay the stopping
rule on the exact energies of the same sessions' trajectories. Each
trajectory keeps its initial jitter and its shot-noisy gradient steps,
but the stopping decision no longer sees readout noise. The spread of
this ``process'' stopping time is the part of the within-class spread
that the optimizer path creates. Table~\ref{tab:noise} shows the
decomposition. The process standard deviation is almost independent of
the shot budget (\RVprocSdLo{} to \RVprocSdHi{} iterations), so it is
dominated by a shot-independent source, the initial jitter, and its
share of the within-class variance rises from \RVprocLo{} at $256$
shots to \RVprocHi{} at $16384$. This floor explains why the leakage levels off at large shot
budgets. The skewness of the measured laws is
positive in every instance (\RVskewLo{} to \RVskewHi{}), the process component is even more right-skewed, and the readout contribution
pulls the skew down, as Proposition~\ref{prop:ig} and the negative skew
of~\eqref{eq:exactlaw} together predict.

\begin{figure}[!htbp]
\centering
\includegraphics[width=0.62\textwidth]{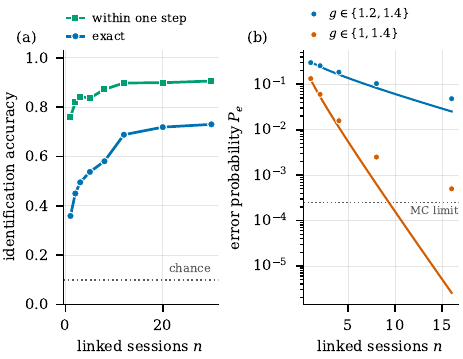}
\caption{Instance identification from $n$ linked sessions of one
client. (a) Ten-instance accuracy of a held-out Bayes adversary, scored
exactly and to within one grid step. (b) Two-instance error
probability: markers are held-out measurements from $4000$ trials,
lines the closed form~\eqref{eq:lecam} with $\Delta t$ and $\sigma$
from the training half. The dotted line marks the Monte Carlo
resolution; points with no error are omitted.}
\label{fig:sessions}
\end{figure}

\subsection{Cross-session accumulation}
Figure~\ref{fig:sessions}(a) gives the accuracy of a held-out adversary
given $n$ linked sessions of one client, drawn from $50$ held-out
sessions per instance. By $n=\NUMnsessionsMax$ it names the exact
instance \NUMaccMax{} of the time and lands within one step
\NUMaccTolMax{} of the time. The residual error sits on adjacent fields
in the disordered phase, whose stopping-time laws overlap
(Fig.~\ref{fig:traj}(b)). In other words, the round count reliably reveals
which region of the phase diagram the client is working in, but not
always the exact field.

Figure~\ref{fig:sessions}(b) tests Hypothesis~\ref{hyp:lecam}. For each
pair, $\Delta t$ and $\sigma$ are estimated on one half of the sessions,
and the Bayes test, with likelihoods fitted on that half, is scored on
the other. For the hardest pair shown, $g\in\{1.2,1.4\}$, the training
half gives $\Delta t=\RVtpdt$ and $\sigma=\RVtpsig$ iterations,
so~\eqref{eq:nsessions} predicts $\RVtpN$ sessions for $5\%$ error.
The held-out error at $n=16$ is \RVtpEmpSixteen{}, against
\RVtpThSixteen{} predicted. Across the pairs and all $n\le16$ the
closed form is within \RVtpMaxDiff{} percentage points of the
measurement, but it underestimates the error in the tail, where the
right-skewed laws of Section~\ref{subsec:noise} depart from the
Gaussian. Equation~\eqref{eq:nsessions} gives a useful order-of-magnitude estimate, not a guarantee.

\begin{figure}[!htbp]
\centering
\includegraphics[width=0.62\textwidth]{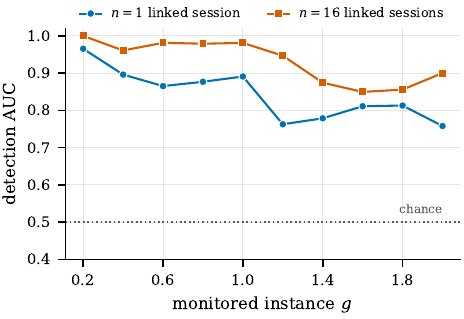}
\caption{Open-world detection. Each point takes one instance as
monitored and the other nine as background; the detection AUC of a
held-out likelihood-ratio detector is shown for one and for sixteen
linked sessions.}
\label{fig:openworld}
\end{figure}

\subsection{Open-world detection}
\label{subsec:openworld}
Work on website fingerprinting showed that closed-world accuracy
overstates an attack, and that a fair evaluation places a monitored
target among unmonitored ones~\cite{dyer_peekaboo_2012}. We designate each instance in turn as
monitored; the server sees sessions drawn either from it or from the
other nine and must decide which, using a likelihood-ratio detector
fitted on held-out data. Figure~\ref{fig:openworld} gives the area
under the ROC curve (AUC) for every choice of monitored instance. From
one session it ranges from \RVowOneLo{} to \RVowOneHi{} (median
\RVowOneMed{}), lowest for $g=\RVowWorstG$, and with $16$ linked
sessions from \RVowSixLo{} to \RVowSixHi{} (median \RVowSixMed{}). The
strict operating points are weaker. At a false-positive rate of $1\%$
and $n=16$, the median detection rate is \RVtprMed{}, ranging from
\RVtprLo{} to \RVtprHi{} across monitored instances. Taken together, the round count is a strong signal in aggregate, but for
most targets it is weak evidence about any single session. Our background consists of other instances of the same
family; a provider's real background of unrelated workloads would
change the base rates in ways this experiment cannot capture.

\section{Generality}
\label{sec:generality}

\begin{table}[!htbp]
\centering
\caption{Leakage across configurations. $|\Theta|$ is the number of
equiprobable instances and $\SNR_{\mathrm{eff}}=\sqrt{\kappa-1}$
from~\eqref{eq:ub}. ``Full sample'' gives the bias-corrected leakage and
bound~\eqref{eq:ub} on all sessions; ``Cross-fitted'' gives the bound from half A
and the leakage on half B, averaged over $40$ splits. Entries marked
$\dagger$ have permutation $p>0.05$; all others have $p\le0.005$.
Ising runs use $1024$ shots unless a shot count is given.}
\label{tab:systems}
\footnotesize
\setlength{\tabcolsep}{4pt}
\begin{tabular}{@{}lcccccc@{}}
\toprule
& & & \multicolumn{2}{c}{Full sample} & \multicolumn{2}{c}{Cross-fitted} \\
\cmidrule(lr){4-5}\cmidrule(l){6-7}
Configuration & $|\Theta|$ & $\mathrm{SNR}_{\rm eff}$ & $I$ & Bound & Pred. & Meas. \\
\midrule
GD 256 & 10 & 1.90 & $0.830$ & 1.104 & 1.102 & $0.705$ \\
GD 1024 & 10 & 2.74 & $1.236$ & 1.545 & 1.548 & $1.160$ \\
GD 4096 & 10 & 3.81 & $1.463$ & 1.976 & 1.980 & $1.283$ \\
GD 16384 & 10 & 4.70 & $1.684$ & 2.265 & 2.272 & $1.492$ \\
indep. init & 10 & 1.02 & $0.431$ & 0.516 & 0.543 & $0.371$ \\
$n=4$ & 10 & 1.11 & $0.528$ & 0.577 & 0.595 & $0.470$ \\
$n=8$ & 10 & 2.04 & $0.807$ & 1.185 & 1.200 & $0.693$ \\
SPSA & 10 & 1.04 & $0.295$ & 0.529 & 0.532 & $0.209$ \\
Max-Cut SPSA$^{\dagger}$ & 6 & 0.21 & $0.007$ & 0.030 & 0.050 & $-0.006$ \\
Max-Cut GD & 6 & 0.51 & $0.141$ & 0.168 & 0.176 & $0.128$ \\
Max-Cut graph$^{\dagger}$ & 20 & 0.20 & $-0.002$ & 0.029 & 0.057 & $0.002$ \\
\bottomrule
\end{tabular}

\end{table}

We now change, one at a time, the parts of the setup that could be
responsible for the channel. Table~\ref{tab:systems}
collects the results.

\subsection{Optimizer}
Replacing parameter-shift gradient descent by simultaneous perturbation
stochastic approximation (SPSA)~\cite{spall_multivariate_1992}, a common choice on hardware, gives mean stopping times from
\NUMspsaTlo{} to \NUMspsaThi{} iterations across the family and a
leakage of \RVspsaI{}~bits ($p\le0.005$). SPSA's random search
directions widen the conditional laws, and $\SNR_{\mathrm{eff}}$ falls
from \RVmainSNR{} to \RVspsaSNR{}.

\subsection{Initialization}
With the common initial point replaced by an independent draw
$\theta_0\sim\mathcal N(0,0.35^2)$ per component for every session
($100$ sessions per instance, $1024$ shots, $T_{\max}=200$), the leakage
falls from \RVmainI{} to \RVindI{}~bits (held-out lower bound
\RVindBA{}) and single-session accuracy from \RVaccMain{} to
\RVindAcc{}, with $\SNR_{\mathrm{eff}}=\RVindSNR$. Spreading the initial points is a cheap and effective
protection, but it is not sufficient on its own. The main configuration
represents a client that reuses its initialization.

\subsection{System size}
At $n=4$ and $n=8$ sites (same ansatz depth; $100$ and $60$ sessions per
instance; tolerance scaled with the extensive energy as
$\eps_n=0.01\,n/6$, so that $n=6$ reproduces the main run) the leakage
is \RVnFourI{} and \RVnEightI{}~bits, against \RVmainI{} at $n=6$. The
channel persists at all three sizes. The dependence on size is not
monotone, and three sizes at simulable scale cannot establish a trend;
Section~\ref{sec:discussion} returns to this.

\subsection{Problem family}
For QAOA at depth $p=3$ on Max-Cut instances with $8$ vertices, the
secret is the edge-density class (random graphs with $8$ to $18$ edges
in steps of two; $25$ graphs per class, $6$ restarts each). Under SPSA
the leakage is \RVqsI{}~bits with $p=\RVqsP$, not distinguishable from
zero. That result alone would confound family with optimizer, so we
repeated it with gradient descent on central finite-difference
gradients of shot-noisy costs (step $0.1$, learning rate $0.05$,
tolerance $0.002$, $w=10$): the leakage is then \RVqgI{}~bits with
$p=\RVqgP$, below its bound of \RVqgBound{}. Asking which of $20$
graphs with $12$ edges the client solves, under SPSA, gives
$\RVqpI$~bits with $p=\RVqpP$. For Max-Cut the channel is weak: it is closed under SPSA and
open under gradient descent. Whether a problem leaks depends on the
optimizer as well as on the problem family.

\begin{table}[!htbp]
\centering
\caption{Leakage under four stopping rules applied to the same recorded
sessions at $1024$ shots. ``Cross-fit'' gives predicted and measured
values as in Table~\ref{tab:systems}; the first row is the GD 1024 configuration of that table, cross-fitted over independent random splits. All permutation $p\le0.005$.}
\label{tab:rules}
\footnotesize
\setlength{\tabcolsep}{4pt}
\begin{tabular}{@{}lcccc@{}}
\toprule
Stopping rule & $\Ex[T]$ & $I(T;g)$ & Bound & Cross-fit \\
\midrule
two-window slope, $\varepsilon=0.01$ & 44 & 1.236 & 1.545 & 1.546 / 1.159 \\
two-window slope, $\varepsilon=0.005$ & 50 & 0.914 & 1.131 & 1.135 / 0.837 \\
relative change, $10^{-3}$ & 62 & 0.392 & 0.402 & 0.408 / 0.342 \\
patience, 15 iterations & 81 & 0.208 & 0.295 & 0.295 / 0.169 \\
\bottomrule
\end{tabular}

\end{table}

\subsection{Stopping rule}
Because the cost histories are recorded, other stopping rules can be
applied to the same sessions exactly: a tighter tolerance, a
relative-change criterion, and patience-based early stopping, which
halts when the running minimum of the smoothed cost has not improved
for $15$ iterations. All four rules leak (Table~\ref{tab:rules}), from
$1.24$ down to $0.21$~bits, so the channel is a property of adaptive
stopping rather than of one criterion. The relative-change rule sits
closest to its bound, at $0.392$ bits measured against $0.402$ on the
full sample. If $\kappa$ is approximated by a simple ratio of between-instance to
within-instance variance, this rule appears to exceed the bound; with
the exact form of~\eqref{eq:ub} it does not.

\begin{figure}[!htbp]
\centering
\includegraphics[width=0.62\textwidth]{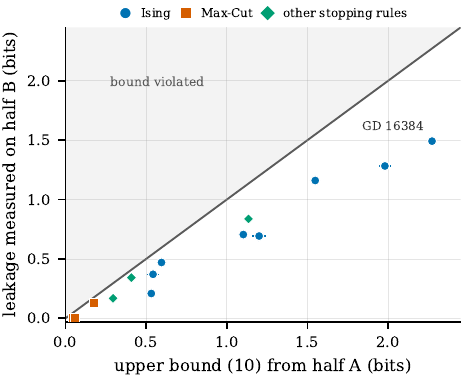}
\caption{Cross-fitted test of the second-moment envelope
(Hypothesis~\ref{hyp:snr}). Each point is one configuration of
Tables~\ref{tab:systems} and~\ref{tab:rules}: bound~\eqref{eq:ub}
estimated on half A of the sessions against the leakage measured on
half B, averaged over $40$ splits, with one-standard-deviation bars.
Negative estimates are drawn at zero. Points in the shaded region would
violate the bound.}
\label{fig:universal}
\end{figure}

\subsection{The second-moment envelope, cross-fitted}
\label{subsec:xfit}
Table~\ref{tab:systems}, Table~\ref{tab:rules} and
Fig.~\ref{fig:universal} test Hypothesis~\ref{hyp:snr}. Averaged over
splits, the leakage measured on half B lies below the bound predicted
from half A in all \RVnCfgU{} configurations, and in each configuration
it does so in at least \RVxfWorst{} of individual splits; the lowest
rate, for the \RVxfWorstName{}, occurs where the bound and the
measurement are within one standard deviation of each other. For the
\RVnSigU{} configurations with significant leakage the ratio of
measured to predicted lies between \RVxfRatioLo{} and \RVxfRatioHi{}.
In practice the bound is a useful screening quantity, since a client
can compute it from the means and variances of its own stopping times.
It is not an accurate predictor, and by Remark~\ref{rem:counter} it is
not a certificate either: stopping-time distributions that are far from
Gaussian can leak more than the bound suggests.

\begin{table}[!htbp]
\centering
\caption{Held-out accuracy of a Bayes adversary trained on the row
configuration and tested on the column configuration (ten instances,
chance $0.10$). Diagonal entries use disjoint halves of the same
configuration.}
\label{tab:transfer}
\footnotesize
\setlength{\tabcolsep}{4pt}
\begin{tabular}{@{}lcccc@{}}
\toprule
\multicolumn{5}{@{}l}{(a) train on shot budget (rows), test on (columns)} \\
& 256 & 1024 & 4096 & 16384 \\
\midrule
256 & \textbf{0.24} & 0.30 & 0.33 & 0.32 \\
1024 & 0.26 & \textbf{0.34} & 0.41 & 0.47 \\
4096 & 0.24 & 0.34 & \textbf{0.48} & 0.53 \\
16384 & 0.24 & 0.35 & 0.45 & \textbf{0.51} \\
\midrule
\multicolumn{5}{@{}l}{(b) train on stopping rule (rows), test on (columns)} \\
& slope 0.01 & slope 0.005 & relative & patience \\
\midrule
slope 0.01 & \textbf{0.33} & 0.14 & 0.07 & 0.06 \\
slope 0.005 & 0.13 & \textbf{0.25} & 0.10 & 0.05 \\
relative & 0.01 & 0.15 & \textbf{0.17} & 0.14 \\
patience & 0.01 & 0.03 & 0.13 & \textbf{0.14} \\
\bottomrule
\end{tabular}

\end{table}

\subsection{What the adversary must know}
\label{subsec:transfer}
The threat model grants the server the client's configuration.
Table~\ref{tab:transfer} measures the cost of getting it wrong. An
adversary trained at one shot budget and applied at another keeps
accuracy between \RVxsOffLo{} and \RVxsOffHi{}, against \RVxsDiag{} for
a matched adversary at $1024$ shots and $0.10$ for chance; accuracy on
the test configuration is limited mainly by how separable that
configuration is, not by the mismatch, because the stopping-time
locations move little with the shot budget. An adversary trained on the
wrong stopping rule, by contrast, falls to a mean accuracy of
\RVxrOffMean{} (at most \RVxrOffMax{}), close to chance, because
different rules place the stopping times of the same instance at
different rounds. The attack thus needs the client's stopping rule. This offers
limited protection, because stopping rules are often library defaults,
but it suggests that the stopping rule is the part of the configuration
most worth keeping private.

\subsection{Quality of the Gaussian approximation}
\label{subsec:gof}
The Gaussian approximation~\eqref{eq:surrogate} enters only the closed forms; every
leakage figure is non-parametric. On the main configuration the
Lilliefors-type test of Section~\ref{subsec:stats} rejects normality at
the $5\%$ level in \RVlfReject{} of the $10$ instances without
correction for multiplicity, and in none after a Holm correction. The
skewness is positive throughout (\RVskewLo{} to \RVskewHi{}), with
bootstrap intervals excluding zero in nine of ten instances, as
Section~\ref{subsec:noise} explains. The Gaussian approximation is adequate for the scaling
statements, but it misses the right tail, which is where
Theorem~\ref{thm:lb} underestimates the error.

\section{Mitigation}
\label{sec:mitigation}

The structural analog of this problem in classical privacy is an
analyst who decides adaptively, from private data, how many queries to
run: the stopping decision leaks even when every answer is private.
Rogers, Roth, Ullman and Vadhan formalized this with privacy filters
and odometers~\cite{rogers_privacy_2016}, with more recent
realization-level refinements~\cite{taylor_realisation-level_2026}.
In our setting it is the client's own stopping rule that leaks, but the
same tools apply. We first show what defenses acting after the stopping
decision cannot achieve, and then give two defenses that act before it,
each with a guarantee and a measured cost.

\begin{definition}[$\Delta$-adjacency]
\label{def:adjacency}
Two executions with realized stopping times $T,T'\in\Nat$ are
\emph{$\Delta$-adjacent} if $|T-T'|\le\Delta$. A release mechanism $M$
is $(\eps,\delta)$-differentially private at adjacency $\Delta$ if
$\Pr[M(T)\in S]\le e^{\eps}\Pr[M(T')\in S]+\delta$ for all
$\Delta$-adjacent $T,T'$ and all $S$.
\end{definition}

\subsection{Defenses after the stop can only pad}
\begin{lemma}[Defenses after the stop can only pad]
\label{lem:posthoc}
Let the client's rule fire at $\tau$, a stopping time with respect to
its own filtration, and let $M$ be any mechanism whose only control is
over the delegation schedule from the moment the rule fires. Then
$M(\tau)\ge\tau$ almost surely.
\end{lemma}

\begin{proof}
The $\tau$ iterations have been submitted and served by the time the
rule fires, and the server's count of served rounds is non-decreasing
in time. No later action can reduce it.
\end{proof}

\begin{theorem}[Padding-only mechanisms are expensive]
\label{thm:impossible}
Let $M$ be any mechanism with $M(T)\ge T$ almost surely. If $M$ is
$(\eps,\delta)$-differentially private at adjacency $\Delta$, then
\begin{equation}
 \Ex\big[M(T)-T\big]\;\ge\;\Delta\,(1-\delta)
 \label{eq:impossible}
\end{equation}
for every $\eps$.
\end{theorem}

\begin{proof}
Fix $T$ and let $T'=T+\Delta$, which is $\Delta$-adjacent to $T$. Let
$S=\{y: y<T+\Delta\}$. Since $M(T')\ge T+\Delta$ almost surely,
$\Pr[M(T')\in S]=0$, so the privacy condition gives
$\Pr[M(T)\in S]\le\delta$. Writing $P=M(T)-T\ge0$ for the padding, this
says $\Pr[P<\Delta]\le\delta$, hence
$\Ex[P]\ge\Delta\Pr[P\ge\Delta]\ge\Delta(1-\delta)$.
\end{proof}

Bound~\eqref{eq:impossible} does not involve $\eps$. As long as $\delta$ is small, a
padding-only scheme pays almost the full adjacency radius on average,
for any privacy level. Cheaper random padding, such as one-sided
geometric padding, has no useful differential-privacy guarantee and,
as the measurements in Section~\ref{subsec:tradeoffs} show, reduces the
measured leakage only
slowly. A cheaper defense must therefore act before the stopping
decision.

\subsection{A fixed public budget}
\begin{proposition}[Fixed public budget]
\label{prop:budget}
Let $B$ be fixed before the session, independently of the instance. The
client runs its own rule, which fires at $\tau_0$; it executes
$\min(\tau_0,B)$ real iterations, then $(B-\tau_0)^{+}$ dummy iterations
that delegate circuits of the same shape with fresh pads, and returns
the iterate held at $\min(\tau_0,B)$. Then the server's record has the
same law for every instance, so
$\Iinfo(\mathcal O;\Theta)=0$. The client executes exactly $B$ iterations, and \rev{the expected
change in solution quality,
$\Delta E_B=\Ex\big[E_{\min(\tau_0,B)}-E_{\tau_0}\big]$, where $E_t$
is the exact energy of the iterate held at round $t$, vanishes when
$B\ge\tau_0$ almost surely. If, in addition, the exact energy is
non-increasing along every trajectory from round $B$ onward, then
$\Delta E_B\ge0$.}
\end{proposition}

\begin{proof}
The released count is the constant $B$. By Lemma~\ref{lem:pad} the
transcript of every round, real or dummy, has law $\mu_1$ whatever
circuit is delegated, so Corollary~\ref{cor:fixedT} applies. \rev{On
the event $\{\tau_0\le B\}$ the returned iterate is the one at the
natural stop, so the difference inside the expectation is zero; if
$B\ge\tau_0$ almost surely, this is the only event and
$\Delta E_B=0$. On $\{\tau_0>B\}$ the difference is
$E_B-E_{\tau_0}$, which is non-negative under the monotonicity
assumption, so $\Delta E_B\ge0$.}
\end{proof}

\begin{revised}
Without the monotonicity assumption, $\Delta E_B$ can be negative. A
stochastic optimizer does not decrease the energy at every step:
finite-shot noise in the gradient estimates, or a gradient-free method
such as SPSA, can move the parameters uphill, so an earlier iterate
can have a lower energy than the one held at the natural stop. The
fixed budget then returns a better iterate than the unprotected run.
This does not affect the privacy statement, which concerns only the
released count. In our gradient-descent simulations the assumption holds in nearly all
sessions: the exact energy rises at some iteration
before the natural stop in \RRriseMain{} of sessions with a common
initialization and in \RRriseInd{} with independent initializations,
and none of the \RRtruncMain{} and \RRtruncInd{} truncated sessions,
over all budgets considered in Section~\ref{subsec:tradeoffs}, returned an iterate
with lower exact energy than the one at its natural stop.
\end{revised}

The only requirement is that $B$ must not depend on the instance. It
should be set from information about the whole problem class, such as pilot runs over
the problem class or a published benchmark, and not from the client's
experience with the instance at hand. A budget below a run's natural stopping round cuts that run short.
This is the price of removing the channel without paying for the
longest run in every session.

\subsection{Differentially private stopping}
A budget gives up adaptivity. To keep an adaptive rule with a formal
guarantee, the client can make the stopping decision itself
differentially private with the AboveThreshold algorithm, the core of
the sparse vector
technique~\cite{dwork_complexity_2009,dwork_algorithmic_2014,lyu_understanding_2017}.
Algorithm~\ref{alg:svt} applies it to the slope
statistic~\eqref{eq:rule}, clipped to a band of width $\beta$ around
the log-tolerance.

\begin{algorithm}[!htb]
\caption{AboveThreshold stopping $M_\eps$}
\label{alg:svt}
\begin{algorithmic}[1]
\STATE \textbf{Input:} privacy budget $\eps$, band $\beta$, tolerance
 $\eps_{\mathrm{stop}}$, public $t_{\min}$ and $T_{\mathrm{cap}}$
\STATE $\tau\leftarrow-\ln\eps_{\mathrm{stop}}$; draw
 $\rho\sim\Lap(2\beta/\eps)$
\FOR{$t=t_{\min},\dots,T_{\mathrm{cap}}-1$}
\STATE compute $S_t$ from~\eqref{eq:rule} on the client's own
 decrypted costs
\STATE $f_t\leftarrow\mathrm{clip}\big(-\ln S_t,\ \tau-\beta/2,\ \tau+\beta/2\big)$
\STATE draw $\nu_t\sim\Lap(4\beta/\eps)$
\IF{$f_t+\nu_t\ge\tau+\rho$}
\STATE halt and return the iterate at $t$ \COMMENT{released count $R=t$}
\ENDIF
\ENDFOR
\STATE halt at $R=T_{\mathrm{cap}}$
\end{algorithmic}
\end{algorithm}

\begin{theorem}[Instance-level privacy of $M_\eps$]
\label{thm:svt}
For any two instances $\theta,\theta'$, with any laws of their cost
histories, and any set $S$ of rounds,
$\Pr[R\in S\mid\theta]\le e^{\eps}\,\Pr[R\in S\mid\theta']$. The same
holds for the server's whole record, and
$\Iinfo(\mathcal O;\Theta)\le\eps\log_2 e$ bits for any prior.
\end{theorem}

\begin{proof}
Fix two cost histories $h,h'$. Every clipped query satisfies
$|f_t(h)-f_t(h')|\le\beta$, because both values lie in an interval of
width $\beta$. AboveThreshold with threshold noise $\Lap(2\beta/\eps)$
and query noise $\Lap(4\beta/\eps)$, stopped at the first query above
the threshold, is $\eps$-differentially private for any two inputs
whose query answers differ by at most $\beta$~\cite[Thm.~3.23]{dwork_algorithmic_2014};
halting at $T_{\mathrm{cap}}$ is the output reporting that no query was
above. Hence $\Pr[M(h)\in S]\le e^{\eps}\Pr[M(h')\in S]$ for every
$h,h'$. The history does not depend on the mechanism's noise, since the
optimizer's trajectory up to any round is the same whether or not it is
stopped there; averaging over $h\sim P_\theta$ and then over
$h'\sim P_{\theta'}$ gives the instance-level statement. By
Corollary~\ref{cor:info} the server's record is $R$ plus
instance-independent noise, so post-processing extends the guarantee to
it. Finally, pure $\eps$-differential privacy bounds every log-likelihood
ratio, and hence the mutual information, by $\eps$ nats.
\end{proof}

The guarantee holds for every pair of instances without choosing an
adjacency radius, which is its advantage over
Definition~\ref{def:adjacency}. Its weakness is solution quality.
Early in a run, before convergence, the clipped statistic sits at its
lower end, and the mechanism can still fire.

\begin{proposition}[Premature stopping]
\label{prop:premature}
Suppose that in $k$ rounds $t_{\min}\le t<t_{\min}+k$ the client's
statistic is at the lower clip, $f_t=\tau-\beta/2$, meaning that the
cost is still falling at least $e^{\beta/2}$ times faster than the
tolerance. Then
\begin{equation}
 \Pr[R<t_{\min}+k]\;\ge\;\tfrac12\Big[1-\big(1-\tfrac12e^{-\eps/8}\big)^{k}\Big].
 \label{eq:premature}
\end{equation}
\end{proposition}

\begin{proof}
With probability $\frac12$, $\rho\le0$. On that event the mechanism fires
at round $t$ whenever $\nu_t\ge\beta/2+\rho$, which is implied by
$\nu_t\ge\beta/2$, an event of probability
$\frac12e^{-(\beta/2)/(4\beta/\eps)}=\frac12e^{-\eps/8}$, independently
across rounds and of $\rho$.
\end{proof}

Bound~\eqref{eq:premature} does not depend on $\beta$, so widening the band does not
help. For $k=20$ it gives \RVsvtLBEight{} at $\eps=8$ and
\RVsvtLBSixteen{} at $\eps=16$: at privacy levels where the guarantee is meaningful, a large fraction
of runs are cut off while the optimizer is still making fast
progress.

\begin{figure}[!htbp]
\centering
\includegraphics[width=0.62\textwidth]{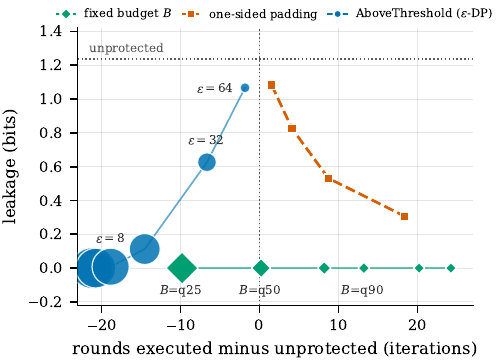}
\caption{Leakage against the change in executed iterations for the
three mechanism families. Marker area is proportional to the mean
energy penalty; padding has none. Budgets are labeled by the quantile
of the unprotected stopping times they equal, AboveThreshold points by
$\eps$.}
\label{fig:mitigation}
\end{figure}

\begin{table}[!htbp]
\centering
\caption{Mitigations replayed on the recorded sessions at $1024$
shots. Leakage is the bias-corrected mutual information in bits (a
constant release has exactly zero); Acc.\ is held-out single-session
accuracy (chance $0.100$); $\Delta$rounds is the mean change in executed
iterations (multiply by \NUMrounds{} for delegated rounds); Early is the
fraction of runs returning before their natural stop; $\Delta E$ is the
mean energy penalty. Budgets are quantiles of the pooled unprotected
stopping times; AboveThreshold uses $\beta=3$, $t_{\min}=20$.}
\label{tab:mitigation}
\footnotesize
\setlength{\tabcolsep}{3pt}
\begin{tabular}{@{}llccccc@{}}
\toprule
Mechanism & Setting & Leakage & Acc. & $\Delta$rounds & Early & $\Delta E$ \\
\midrule
none & n/a & 1.236 & 0.343 & 0.0 & 0\% & 0.000 \\
budget & $B=44$ (q50) & 0 & 0.100 & $+0.2$ & 50\% & 0.032 \\
budget & $B=52$ (q75) & 0 & 0.100 & $+8.2$ & 24\% & 0.007 \\
budget & $B=57$ (q90) & 0 & 0.100 & $+13.2$ & 9\% & 0.001 \\
budget & $B=64$ (q99) & 0 & 0.100 & $+20.2$ & 1\% & 0.000 \\
pad only & $p=0.2$ & 0.826 & 0.251 & $+4.1$ & 0\% & 0.000 \\
pad only & $p=0.05$ & 0.306 & 0.167 & $+18.4$ & 0\% & 0.000 \\
AboveThr. & $\varepsilon=8$ & 0.008 & 0.108 & $-18.8$ & 96\% & 0.338 \\
AboveThr. & $\varepsilon=16$ & 0.113 & 0.137 & $-14.5$ & 92\% & 0.230 \\
AboveThr. & $\varepsilon=32$ & 0.626 & 0.248 & $-6.6$ & 78\% & 0.076 \\
AboveThr. & $\varepsilon=64$ & 1.067 & 0.301 & $-1.8$ & 52\% & 0.011 \\
\bottomrule
\end{tabular}

\end{table}

\subsection{Measured trade-offs}
\label{subsec:tradeoffs}
We evaluated all mechanisms by replaying their stopping decisions on
the recorded sessions of the main configuration. Replay is exact,
because the trajectory up to any round does not depend on when the run
is later stopped. Table~\ref{tab:mitigation} and
Fig.~\ref{fig:mitigation} give the results; the mean unprotected run
length is \RVunprotRounds{} iterations and the mean residual energy gap
at the unprotected stop, relative to the exact ground state, is
\RVresGap{}.

A fixed budget at the median run length ($B=\RVbudFiftyB$) removes the
channel exactly with essentially no change in the mean number of
iterations ($\RVbudFiftyRounds$); it truncates \RVbudFiftyTrunc{} of
runs and costs a mean energy penalty of \RVbudFiftydE{} \rev{(positive
for every truncated session, as noted after
Proposition~\ref{prop:budget})}, which is
\RVbudFiftyPct{} of the residual gap. At the $75$th percentile
($B=\RVbudSevFiveB$) the penalty falls to \RVbudSevFivedE{}
(\RVbudSevFivePct{} of the residual gap), at a cost of
\RVbudSevFiveExtra{} extra iterations per run. With independent initializations the median budget
costs \RVindBudFiftydE{}, so the conclusion does not depend on the
initialization regime. One-sided geometric padding, by comparison,
still leaks \RVpadFiveI{}~bits after adding \RVpadFiveExtra{} iterations
per run, while a budget at the $99$th percentile, which adds
\RVbudNineNineExtra{}, leaks
nothing, as Theorem~\ref{thm:impossible} anticipates.

AboveThreshold stopping behaves as Proposition~\ref{prop:premature}
predicts. At $\eps=64$, where the formal guarantee is vacuous, the
leakage is still \RVsvtSixFourI{}~bits. At $\eps=16$ it falls to
\RVsvtSixteenI{}~bits, but \RVsvtSixteenEarly{} of runs stop before
their natural round and the mean energy penalty is \RVsvtSixteendE{},
\RVsvtSixteenPct{} of the residual gap. At $\eps=8$,
\RVsvtEightEarly{} of runs stop early with a penalty of
\RVsvtEightdE{}. In this setting differentially private stopping is
dominated by a fixed budget, which achieves zero leakage at a small
fraction of the cost. We recommend the budget, and reserve
AboveThreshold for clients that cannot fix a budget in advance and can
tolerate a weaker optimum.

\section{Related work}
\label{sec:related}

\textbf{Blind and verifiable delegation.} BFK~\cite{broadbent_universal_2009}
and FK~\cite{fitzsimons_unconditionally_2017} give the single-round
guarantees; Dunjko \emph{et al.}~\cite{dunjko_composable_2014} make
them composable in the abstract-cryptography
framework~\cite{maurer_abstract_2011,portmann_security_2022}; Leichtle
\emph{et al.}~\cite{leichtle_verifying_2021} and Kapourniotis \emph{et
al.}~\cite{kapourniotis_unifying_2024} reduce the overhead and add
noise robustness; Aharonov \emph{et al.}~\cite{aharonov_interactive_2008}
give a circuit-model alternative. None treats the length of a
delegation sequence as leakage to be quantified.

\textbf{Delegated variational algorithms.} Inajetovic, Wallden and
Pappa~\cite{inajetovic_verifiable_2026} give an end-to-end verifiable
delegated VQA with convergence and soundness guarantees for the loop;
Li \emph{et al.}~\cite{li_delegated_2023} use quantum homomorphic
encryption for the same task. We take such a protocol as given and
analyze what its control flow discloses.

\textbf{Landscape geometry and trainability.} The analysis of
Section~\ref{sec:mechanism} rests on the PL
condition~\cite{polyak_gradient_1963,karimi_linear_2016,nesterov_introductory_2004},
and the quantity it tracks is tied by~\eqref{eq:pl} to the gradient
magnitude whose exponential decay defines barren
plateaus~\cite{larocca_barren_2025}. To our knowledge this quantity
has not previously been considered as a confidentiality risk.

\textbf{Quantum side channels.} Timing
attacks~\cite{lu_quantum_2025} and power-trace
attacks~\cite{xu_exploration_2023,erata_quantum_2024,bell_reconstructing_2022} reconstruct circuit structure by measuring physical signals
from single executions. Our channel is coarser, needs no measurement
apparatus, and concerns the sequence rather than the circuit; the two could be combined.

\textbf{Traffic analysis.} Hintz~\cite{hintz_fingerprinting_2003}
established that packet counts and timing identify encrypted content,
and Dyer \emph{et al.}~\cite{dyer_peekaboo_2012} showed that cheap
padding defenses against it fail, an empirical finding whose structure
Theorem~\ref{thm:impossible} reproduces for one-sided mechanisms. That
literature also supplied the open-world protocol of
Section~\ref{subsec:openworld}. Kocher's timing
attacks~\cite{kocher_timing_1996} are the origin of the constant-time
discipline that the fixed budget instantiates.

\textbf{Privacy of adaptive stopping.} Rogers \emph{et
al.}~\cite{rogers_privacy_2016} formalize data-dependent stopping in
differential privacy, building on the Laplace
mechanism~\cite{dwork_calibrating_2006} and its discrete
counterpart~\cite{ghosh_universally_2012}; see
also~\cite{taylor_realisation-level_2026}. The sparse vector technique
originates with Dwork \emph{et al.}~\cite{dwork_complexity_2009}, and
Lyu, Su and Li~\cite{lyu_understanding_2017} document variants that
fail to be private, which is why Algorithm~\ref{alg:svt} follows the
noise scales of~\cite[Alg.~1]{dwork_algorithmic_2014} exactly.

\section{Discussion and limitations}
\label{sec:discussion}

\emph{Simulation, not deployment.} Our sessions are simulated with
exact statevectors and finite-shot sampling. We model neither device
noise nor the queueing and batching of a real service, and we do not
claim a measurement on commercial hardware. A deployment study,
logging only what an honest provider logs across many customer
sessions, would be the most convincing test of these results; the evaluation methodology of website fingerprinting transfers
almost directly.

\emph{Modeling assumptions.} Equation~\eqref{eq:geom} and the Gaussian
approximation~\eqref{eq:surrogate} enter only the closed forms. The
measurements use a real finite-shot optimizer, and
Sections~\ref{subsec:noise} and~\ref{subsec:gof} record where the
closed forms depart from the data, in particular in the right tail.

\emph{Scope of the landscape result.} The early-window forecast
of Section~\ref{subsec:pl_results} is established for ten instances of
one family, with one noiseless trajectory per instance. It shows that in this family the count reflects the early landscape
geometry. It does not show that the same predictor works for other
families without refitting.

\emph{System size and initialization.} The simulations use $4$ to $8$
qubits, which is what statevector simulation permits at the ensemble
sizes the statistics require. The leakage persists at all three sizes
without a monotone trend. The barren-plateau literature predicts
(Remark~\ref{rem:barren}) that
$\mu_t$ shrinks with width and depth, which would lengthen runs and
could either widen or narrow the separation relative to the noise; we
cannot settle which. Initialization matters substantially
(Section~\ref{sec:generality}), and the main configuration is the less
favorable case for the client.

\emph{Timing and active adversaries.} Inter-round wall-clock
timing plausibly carries more, since the duration of a classical
optimizer step can depend on what it computed, and none of our
mitigations addresses it; a fixed budget with fixed per-round timing
would. Theorem~\ref{thm:tight} concerns an honest-but-curious server. A
server that corrupts rounds to move the stopping time is a stronger
adversary, bounded but not analyzed here (Remark~\ref{rem:composable}).

\emph{Adversary knowledge.} Our adversary knows the configuration and
the instance prior. It survives a wrong shot budget and fails with a
wrong stopping rule (Section~\ref{subsec:transfer}). A provider that
does not know the problem class at all faces a harder, open-world
problem than Section~\ref{subsec:openworld} models.

\section{Conclusion}
\label{sec:conclusion}

Blindness and verifiability for delegated quantum computation are
single-round guarantees. We showed that they extend to a delegation
sequence and that, for an honest-but-curious server and fresh
per-round pads, the extension is tight: the whole residual advantage is
the total-variation distance between stopping-time laws, and the
server's information equals the information in the round count.
Because variational algorithms stop adaptively on private data, that
residual is not zero.

In simulated delegated eigensolvers, the round count reports the
landscape the client is optimizing: a local Polyak--{\L}ojasiewicz ratio
measured early in the run forecasts when the client will stop. The
leakage was significant in \RVnSigU{} of \RVnCfgU{} configurations, up
to \RVmaxI{}~bits, and stayed below the held-out bound~\eqref{eq:ub} in all of them. It is reduced but not removed by
independent initialization, it survives changes of optimizer, rule and
size, and it is weak or absent for Max-Cut. Defenses applied after the stop can only pad, and padding cannot be cheap. A fixed public budget removes the channel
exactly at modest cost, while differentially private stopping gives a
formal instance-level guarantee at a much higher cost in solution
quality.

The lesson is more general than our model. Any repeated-delegation
protocol whose control flow, and not merely its per-round content,
depends on private data has a leakage surface that a per-round
definition cannot see. Closing it requires treating the sequence as an object with its own
leakage function, not only the individual rounds.

\data{\rev{The data that support the findings of this study are
available from the authors upon reasonable request.}}

\bibliographystyle{iopartnumdoi}
\bibliography{stopping_time_leakage}

\end{document}